%Paper: cond-mat/9405083
%From: ybkim@MIT.EDU
%Date: Thu, 26 May 94 17:15:12 EDT
%Date (revised): Thu, 22 Sep 94 10:51:37 EDT
%Date (revised): Tue, 15 Nov 1994 23:41:21 EST

%%%%%%%%%%%%%%%%%%%%%%%%%%%%%%%%%%%%%%%%%%%%%%%%%%%%%%%%%%%
%
%  This manuscript is written in plain tex.
%
%%%%%%%%%%%%%%%%%%%%%%%%%%%%%%%%%%%%%%%%%%%%%%%%%%%%%%%%%%%
\magnification=\magstep 1
\baselineskip=20 pt
\font\bigbf = cmbx10 scaled\magstep 1
 1
 1

%\line{}
\line{\hfill\hfill\hfill REVISED VERSION cond-mat/9405083}
\vskip 1.0cm

\noindent

\centerline{\bigbf \hfill Gauge-invariant response functions of \hfill}
\centerline{\bigbf \hfill fermions coupled to a gauge field \hfill}

\vskip 1.0cm

\centerline{\hfill Yong Baek Kim, Akira Furusaki$^*$, Xiao-Gang Wen, and
Patrick A. Lee \hfill}
\vskip 0.3cm
\centerline{\hfill \it Department of Physics, Massachusetts Institute of
Technology \hfill}
\centerline{\hfill \it Cambridge, Massachusetts 02139 \hfill}

\vskip 1.0cm

\centerline{\hfill ABSTRACT \hfill}

\midinsert
\narrower
{\noindent\tenrm

We study a model of fermions interacting with a gauge field and
calculate gauge-invariant two-particle Green's functions or response
functions.
The leading singular contributions from the self-energy correction are
found to be cancelled by those from the vertex correction for small
$q$ and $\Omega$.
As a result, the remaining contributions are not singular enough to
change the leading order results of the random phase approximation.
It is also shown that the gauge field propagator is not
renormalized up to two-loop order.
We examine the resulting gauge-invariant two-particle Green's functions
for small $q$ and $\Omega$, but for all ratios of $\Omega / v_F q$ and
we conclude that they can be described by Fermi liquid forms without any
diverging effective mass.

\vskip 0.2cm
\noindent
PACS numbers: 71.27.+a, 73.40.Hm, 11.15.-q
}
\endinsert

\vfill\vfill\vfill
\break

\vskip 0.5cm

\centerline{\bf I. INTRODUCTION}

The problem of two dimensional fermions coupled to a gauge field has been
a recent subject of intensive research.
This problem appears as a low energy effective model of two different
strongly correlated electronic systems, {\it i.e.}, electrons in the
fractional quantum Hall (FQH) regime and the high-temperature superconductors
(HTSC), both of which have been considered as one of the most important
problems in modern condensed matter physics.

As the first example, this problem arises in a theory of the half-filled Landau
level (HFLL) [1-3] in connection with the composite fermion theory of the
FQH effect [4].
A composite fermion is generated by attaching even number of flux
quanta to an electron [4].
The transformation from the electron to the composite fermion can be
realized by introducing an appropriate Chern-Simons gauge field [1,5].
Especially, at the filling fraction $\nu = 1/2$,
composite fermions see effectively zero magnetic field at the mean field
level [1-4] because of the cancellation between the average of the Chern-Simons
gauge field (from the attached magnetic flux quanta) and the external magnetic
field. Thus, at the mean field level the system can be described as a Fermi
liquid of composite fermions. The fluctuation of the gauge field beyond the
mean field level has been studied within the random phase
approximation (RPA) [1,3], which explains qualitative features of the recent
experiments [6-11].

The other source comes from the recent gauge theory of the normal state
of high temperature superconductors [12-15].
The gauge field arises as a fluctuation of the spin chirality [12] above the
uniform resonating-valence-bond mean field state of the $t-J$ model which is
supposed to be an effective model of HTSC.
The origin of the gauge field fluctuation can be traced back to the
constraint that the doubly-occupied sites are not allowed because of the strong
on-site Coulomb repulsion [12,13].
It has been suggested that the gauge field fluctuations play important
roles in explaining anomalous transport properties of the normal state
of HTSC [12,15,16].

Besides these real examples, the problem of fermions interacting with a gauge
field has been studied as a potential example of non-Fermi liquids [17-28].
In contrast to the usual long-range interactions such as the Coulomb
interaction, the transverse part of the gauge field cannot be screened because
the gauge invariance requires the gauge field to be massless in the absence
of symmetry breaking [17-19].
Thus, one can expect that the long-range interaction due to the transverse
part of the gauge field gives rise to non-Fermi-liquid-like behaviors.
In fact, some singular behaviors appear in the lowest order self-energy
correction of fermions by the gauge field fluctuation [14,17-20].
The singular self-energy correction makes perturbative calculation
unreliable at low energies.
This motivated several non-perturbative calculations of one-particle
Green's function of fermions which show highly non-Fermi-liquid-like
behaviors [21-24,26].
It turns out that, even in the lowest order, the singular self energy
correction
makes the effective mass of the fermion divergent so that
the usual single particle picture breaks down [1].

However, recent experiments on the electrons in the half-filled Landau level
showed essentially Fermi-liquid-like behaviors [6-11] and also measured finite
effective mass of composite fermions [10].
Therefore, we are in a situation
that experiments apparently contradict to the insight we got from the
one-particle Green's function of the fermions.
However, the one-particle Green's function for the fermions is not
gauge invariant.
The singular self-energy correction in the one-particle
Green's function (which leads to divergent effective mass [1]) may be
an artifact of the gauge choice rather than a property of physical
quasi-particles.
Since it is not a gauge-invariant quantity, the one-particle Green's
function for the fermions cannot be directly measured in experiments.
It is possible that some singularities in the gauge-dependent one-particle
Green's function simply do not appear in gauge-invariant correlation
functions.
One purpose of this paper is to examine some gauge-invariant response
functions in order to determine whether the singular behaviors in the
one-particle Green's function appear in gauge-invariant correlation
functions or not.

The importance of the gauge-invariance in calculating correlation
functions can be also seen in the following example.
The leading order corrections (two-loop order) to the transverse
polarization function (or current-current correlation function)
are given by the diagrams in Fig.3.
Note that the sum of contributions from Fig.3 (a)-(d) is not
gauge-invariant because they contain only self-energy corrections but
do not contain the vertex correction.
For concreteness, let us consider the case of $\eta = 2$ in the
model given by Eq.(8), which corresponds to the case of HTSC and
the short-range interaction between fermions in HFLL.
We also consider $\Omega \ll v_F q$ and $q \ll k_F$ limits.
In this case, it can be shown that the correction to the
transverse polarization function due to the self-energy corrections
(given by Fig.3 (a)-(d)) has the following form:
$$
\delta \ {\rm Im} \ \Pi^{s}_{11} ({\bf q}, \Omega) \approx
{m^2 v^3_F \over 2 \pi \gamma} \ {\Omega \over v_F q} \
{(\gamma \Omega / \chi)^{2/3} \over k_F q} \ ,
\eqno{(1)}
$$
while the contribution from the free fermions is given by
$$
{\rm Im} \ \Pi^{0}_{11} ({\bf q}, \Omega) =
- {m v^2_F \over 2 \pi} \ {\Omega \over v_F q} \ ,
\eqno{(2)}
$$
where $1$ denotes the direction which is perpendicular to {\bf q}.
One can see that the correction $\delta \ {\rm Im} \ \Pi^{s}_{11}$
would be more singular than the free fermion contribution
${\rm Im} \ \Pi^{0}_{11}$ if $q, \ \Omega \rightarrow 0$ limit was
taken with fixed $\Omega / v_F q < 1$.
This result suggests that the perturbative expansion breaks down
at low energies and the Fermi-liquid criterion are violated.
Thus the gauge-dependent correction (which comes from the self-energy
correction) to the transverse polarization function provides
a similar picture as that from the singular one-particle Green's
function [29].

Nevertheless, the perturbative corrections to the correlation
functions should be calculated in a gauge-invariant way, thus
one has to include the contributions from the vertex
correction.
The contribution to the transverse polarization function
$\delta \ {\rm Im} \ \Pi^{v}_{11}$ coming from the vertex
correction contains a singular term which exactly cancels
the singular contribution from the self-energy correction.
Thus the remnant terms in $\delta \ {\rm Im} \ \Pi^{v}_{11}$
provide the lowest order corrections to the transverse
polarization function and have the following form:
$$
\delta \ {\rm Im} \ \Pi^{s}_{11} + \delta \ {\rm Im} \ \Pi^{v}_{11}
\approx {m^2 v^3_F \over \gamma} \ {\Omega \over v_F q} \
\left [ \ a \ {(\gamma \Omega / \chi)^{2/3} \over k^2_F} +
b \ {(\gamma \Omega / \chi) \over k^2_F q} \ \right ] \ ,
\eqno{(3)}
$$
where $a, b$ are dimensionless constants.
One can see that the corrections calculated in a gauge-invariant
way are always much less than the free fermion contribution as far as
$\Omega \ll v_F q$ and $q \ll k_F$ limits are concerned.
Therefore, the perturbative expansion works quite well in this regime,
at least up to the leading order gauge field corrections, and
there is no need to go beyond the perturbation theory at this order.
The validity of the perturbative expansion also indicates that the
transverse polarization function is well described by the Fermi-liquid
theory in the region of $\Omega \ll v_F q$ and $q \ll k_F$.
This provides a very different picture from that obtained through the
gauge-dependent one-particle Green's function.

In this paper, we examine several gauge-invariant
two-particle Green's functions or response functions
in the limit of low frequency and long wavelength.
It is shown that all the leading singular contributions from the self-energy
correction are cancelled by the contributions from the vertex correction
in systematic perturbation theories (which guarantee the gauge-invariance
in each order of the perturbative expansion).
This cancellation is essentially due to the Ward identity.
It is found that singular corrections to the two-particle Green's function
do not appear for all ratios of $\Omega / v_F q$ as far as the limit of low
frequency and long wavelength limit is concerned.
This kind of cancellation was also discussed by Ioffe and Kalmeyer [30]
for a static gauge field.
Recently, Khveshchenko and Stamp[23] performed non-perturbative
calculations of one-particle and two-particle Green's functions using
the so-called eikonal approximation.
Even though they obtained a highly singular one-particle Green's function,
the singularity does not show up in two-particle Green's functions for
small $q$ and $\Omega$ in this approximation.

We also show explicitly that the gauge field propagator is not
renormalized by the fluctuations beyond RPA up to two-loop order.
Non-renormalization of the gauge field propagator was first discussed
by Polchinski [28] in the framework of a self-consistent approach.
In this approach, it is assumed that the dispersion relation of fermions
is given by
$\omega \propto \xi^{3/2}_{\bf k}$ ($\xi_{\bf k} = {k^2 / 2 m} - \mu$)
and that of the gauge field is given by $\Omega \propto i q^3$, which
are the results of one-loop corrections.
Ignoring vertex correction by assuming the existence of a Migdal-type
theorem, he showed that
%that higher order corrections do not change
the assumed one-particle Green's function is self-consistent,
and the polarization function
is given by the same form as that of free fermions
${\rm Im} \ \Pi^0_{11} = - (m v^2_F / 2 \pi) \ (\Omega / v_F q)$ for
$\Omega < \gamma^{1/3} \chi^{2/3} q^{3/2}$.
As a result, the gauge field propagator is not renormalized because
the dispersion relation of the gauge field is given by
$\Omega \propto i q^3$.
%, which is self-consistent with the initial assumption.
However, we would like to remark that his result is quite different
from those obtained in this paper.
%First of all,
One can check that the polarization function in
the self-consistent approach has a
different form compared to that of Fermi liquid for $\Omega >
\gamma^{1/3} \chi^{2/3} q^{3/2}$.
%which is contradict to our result.
However, in our perturbative calculation, the cancellation of anomalous
terms from self-energy
and vertex corrections leads to the result that the polarization functions
have Fermi liquid forms for all $q$ and $\Omega$ as far as both are small.
%This is an important difference between our result and the result of
%the self-consistent argument discussed in [28].

We have made several explicit calculations of two-particle Green's functions.
In particular, we consider a model given by Eq.(8)
with $v({\bf q}) = V_0 / q^{2-\eta}$ ($v({\bf r})
\propto V_0 / r^{\eta}, \ 1 < \eta \leq 2$)
%in the model given by Eq.(8),
which corresponds to the interaction between fermions in the problem of
HFLL.
We will present the non-analytic contributions (due to the gauge field
fluctuations) to the two-particle Green's functions.
The transverse polarization function $\Pi_{11} ({\bf q}, \Omega)$ up to
two-loop order is found to have the following form.
For $\Omega \ll v_F q$, we get
$$
{\rm Im} \ \Pi_{11} ({\bf q}, \Omega) \approx - {m v^2_F \over 2 \pi}
\ {\Omega \over v_F q} \ \left [ \ 1 -
a \ {m v_F \over \gamma} {(\gamma \Omega / \chi)^{2 \over 1 + \eta}
\over k^2_F} -
b \ {m v_F \over \gamma} {(\gamma \Omega / \chi)^{3 \over 1 + \eta}
\over k^2_F q} \ \right ] \ ,
\eqno{(4)}
$$
while for $\Omega \gg v_F q$,
$$
{\rm Im} \ \Pi_{11} ({\bf q}, \Omega) \approx
- {1 + \eta \over 8 \pi^2 (5 + \eta)} {1 \over {\rm sin}
\left ( {2 \pi \over 1 + \eta} \right )} {v_F \over m}
{\gamma^{3 - \eta \over 1 + \eta} \over \chi^{4 \over 1 + \eta}}
\Omega^{3 - \eta \over 1 + \eta} \
\left [ \ 1 + c \ m v^3_F
\left (  {\chi \over \gamma} \right )^{1 \over 1 + \eta}
{q^2 \over \Omega^{2 \eta + 3 \over \eta + 1}} \ \right ] \ ,
\eqno{(5)}
$$
where $a, b, c$ are positive dimensionless constants.

The density-density correlation function $\Pi_{00} ({\bf q}, \Omega)$ is also
calculated.
We have a formula valid for any ratio of $\Omega / v_F q$ as long as $\Omega$
and $q$ are small (see Eq.(70)), but here we just discuss limiting cases.
For $\Omega \ll v_F q$, we have
$$
{\rm Im} \ \Pi_{00} ({\bf q}, \Omega) \approx - {m \over 2 \pi}
\ {\Omega \over v_F q} \ \left [ \ 1 -
{1 + \eta \over 4 \pi (5 + \eta)} {1 \over {\rm cos}
\left ( {\eta - 1 \over \eta + 1} \pi \right )}
{1 \over k_F m}
{\gamma^{3 - \eta \over 1 + \eta} \over \chi^{4 \over 1 + \eta}}
\Omega^{3 - \eta \over 1 + \eta}
\left ( {\Omega \over v_F q} \right )^2
\ \right ] \ .
\eqno{(6)}
$$
On the other hand, for $\Omega \gg v_F q$,
$$
{\rm Im} \ \Pi_{00} ({\bf q}, \Omega) \approx
- {1 + \eta \over 8 \pi^2 (5 + \eta)} {1 \over {\rm sin}
\left ( {2 \pi \over 1 + \eta} \right )} {1 \over k_F}
{\gamma^{3 - \eta \over 1 + \eta} \over \chi^{4 \over 1 + \eta}}
\Omega^{3 - \eta \over 1 + \eta}
\left ( {v_F q \over \Omega} \right )^2 \ .
\eqno{(7)}
$$
Note that ${\rm Im} \ \Pi_{11} (q \rightarrow 0, \Omega) =
{\Omega^2 \over v^2_F q^2}
{\rm Im} \ \Pi_{00} (q \rightarrow 0, \Omega)$
is satisfied as it should be.
Eqs.(4)-(7) are the main results of this paper.

{}From the above gauge-invariant correlation functions, one can see that
\item{1)} The corrections are irrelevant in the small $q$ and
$\Omega$ limit regardless of the way how $q$ and $\Omega$ approach to
zero (for example, $q \rightarrow 0$ limit may be taken first or
$\Omega \rightarrow 0$ first, etc.).
Therefore, {\it non-perturbative calculations are not necessary}.
However, the sub-leading contributions are in general non-analytic
due to the long range nature of the gauge interaction. The
non-analytic sub-leading terms may have some experimental consequences.
%
%Considering the behaviors of $\Pi_{00} ({\bf q}, \Omega)$ only for
%small $q$ and $\Omega$ and $\eta = 2$ case,
%one can also discuss
%
For example, the NMR relaxation rate $1 / T_1$ in the problem
of HTSC can be determined from $\Pi_{00} ({\bf q}, \Omega)$.
At low temperatures we have
$$
{1 \over T_1 T} \ \propto \ \lim_{\Omega \rightarrow T} \
- {1 \over \Omega} \ \sum_{\bf q} \ {\rm Im} \
\Pi_{00} ({\bf q}, \Omega) \ ,
$$
where $\Pi_{00}$ plays the role of spin susceptibility in HTSC.
Eq.(6) implies the following non-analytic correction to
the free fermion result (only contributions from small ${\bf q}$ are
considered)
% as a consequence of the long-range gauge interaction, which is given by
$
{1 \over T_1 T} \ \propto \ 1 - A \ T^{5 + \eta \over 1 + \eta} \ ,
$
where $A$ is a constant and the first term is the result of Fermi
liquid.
Notice that this result is in disagreement with
a result based on a renormalization group approach
obtained in Ref.26, even near $\eta=1$.
For HTSC $\eta = 2$ and
$
{1 \over T_1 T} \ \propto \ 1 - A \ T^{7 / 3} \ .
$
Note that the non-analytic correction is very small so that the
Fermi liquid form is preserved.
\item{2)} $q \rightarrow 0$ limit of the transverse polarization function
indicates that the transport scattering rate $\Gamma_{\rm tr}$ (which
determines the DC conductivity) scales as
$\Gamma_{\rm tr} \propto \Omega^{4 \over 1 + \eta}$ at low frequencies
(see Eq.(45) for more details).
This result can be also obtained from the coefficient of the term which
is proportional to $q^2$ in ${\rm Im} \ \Pi_{00} ({\bf q}, \Omega)$, and
the relation
${\rm Im} \ \Pi_{11} (q \rightarrow 0, \Omega) = {\Omega^2 \over v^2_F q^2}
{\rm Im} \ \Pi_{00} (q \rightarrow 0, \Omega)$.
This result exactly agrees with those obtained by different approaches [12,16].
Note that $\Gamma_{\rm tr} < \Omega$ for $1 < \eta \le 2$.
\item{3)} From Eq.(4), one can see that the gauge field corrections are
smaller than the result of free fermions along the curve
$\Omega \propto q^{1 + \eta}$
which is the dispersion relation of the gauge field.
Therefore, the gauge field propagator is not renormalized.
As mentioned above, non-renormalization of the gauge field propagator
was first discussed in Ref.[28] within a self-consistent argument.
\item{4)} For $\eta \le 2$, the gauge field corrections to the
polarization functions
are less singular than the result of the free fermions for
$\Omega < v_F q$.
In particular, the edge of the particle-hole continuum
in ${\rm Im} \ \Pi_{11}$ and
${\rm Im} \ \Pi_{00}$ still occurs at $\Omega \approx {\tilde v_F} q$,
where ${\tilde v_F}$ is finite and shifted from the bare fermi velocity
as in the usual Fermi liquid theory.
We conclude that the two-particle Green's functions are
consistent with those of a Fermi-liquid with a finite effective mass.
However, a combination of a divergent mass and
divergent Fermi-liquid parameters cannot be ruled out.

The remainder of the paper is organized as the following.
In section II, we introduce the model and review some one-particle
properties.
In section III, the transverse polarization function for $q \rightarrow 0$ case
is calculated. The cancellation of anomalous terms (coming from
the self energy and the vertex correction) up to $(1/N)^0$th order
is explicitly shown (where $N$ is the number of species of fermions).
We also discuss the optical conductivity using the information of the
calculated transverse polarization function.
In section IV, we calculate the transverse polarization function for finite
$q \ll k_F$ case.
It is also argued that the gauge field propagator is not renormalized
up to two-loop order.
In section V, the density-density correlation function is calculated
up to two-loop order for finite $q \ll k_F$.
In section VI, the results are compared to the conventional
Fermi-liquid theory and their implication is discussed.
We conclude this paper in section VII.

\vskip 0.5cm

\centerline{\bf II. THE MODEL AND THE ONE-PARTICLE PROPERTIES}

The model is motivated by the above mentioned two strongly-correlated
electronic systems. It is constructed such that it includes the most
important infrared singular behaviors of the one-particle Green's function.
In this paper, we consider the zero temperature limit and use the
Euclidean space formalism.
The model in the Euclidean space is given by
$$
Z = \int \ D \psi \ D \psi^{*} \ D a_{\mu} \ e^{ - \int d\tau \ d^2 r
\ {\cal L}} \ ,
$$
where
$$
\eqalign{
{\cal L} &= \psi^* (\partial_{0} - ia_0 - \mu) \psi - {1 \over 2m} \psi^*
(\partial_i - ia_i)^2 \psi + ia_0 n_f \cr
&+ {\alpha^2 \over 2} \int d^2 r'
(\nabla \times {\bf a}({\bf r})) \ v({\bf r} - {\bf r'})
\ (\nabla \times {\bf a}({\bf r'})) \ . }
\eqno{(8)}
$$
Here $v({\bf q}) = V_0 / q^{2-\eta}$ ($v({\bf r})
\propto V_0 / r^{\eta}, \ 1 < \eta \leq 2$),
$m$ is the bare mass of the fermion, and
$n_f$ is the average density of fermions.
We choose the Coulomb gauge $\nabla \cdot {\bf a} = 0$.
Note that this model is incomplete for the problem of HFLL because of
the absence of the Chern-Simons term.
However, one may expect that it contains possible
low energy singular behaviors because the most singular contribution to the
one-particle properties comes from the transverse part of the gauge field.
In the problem of HFLL, $\alpha = 1/(2 \pi {\tilde \phi})$ and
${\tilde \phi} = 2$ which is the number of flux quanta attached to the
electron [1].
For the case of HTSC, one can take $\alpha = 0$ [12,13].

After integrating out fermions and including gauge field fluctuations up to
one-loop order (RPA), the effective Lagrangian density of the gauge field is
given by [1,12,13]
$$
{\cal L}_{\rm eff} = {1 \over 2} \int {d^2 q \over (2 \pi)^2} {d\omega \over 2
\pi}
\ a^*_{\mu} ({\bf q}, \omega) \ D^{-1}_{\mu \nu} ({\bf q}, i \omega)
\ a_{\nu} ({\bf q}, \omega) \ ,
\eqno{(9)}
$$
where
$$
D^{-1}_{\mu \nu} =
\pmatrix{ \Pi^0_{00} & 0 \cr
0 & \Pi^0_{11} + \alpha^2 v(q) q^2} \ .
\eqno{(10)}
$$
Here $\mu, \nu = 0, 1$ and $1$ represents the direction that is perpendicular
to ${\bf q}$.
$\Pi^0_{00}$ and $\Pi^0_{11}$
are given by the one-loop diagrams in Fig.1 (a) and (b) respectively.
In the limit of $\omega \ll v_F q$, one can find that [1,12,13]
$$
\eqalign{
\Pi^0_{00} &= - {m \over 2 \pi} \left ( 1 - {|\omega| \over v_F q} \right ) \cr
\Pi^0_{11} &= {2 n \over k_F} {|\omega| \over q} + {q^2 \over 24 \pi m} \cr
&\equiv \gamma {|\omega| \over q} + \chi_0 q^2 \ .
}
\eqno{(11)}
$$
Therefore, the gauge field propagator can be expressed as
$$
\eqalign{
D^{-1}_{00} &= - {m \over 2 \pi} \left ( 1 - {|\omega| \over v_F q} \right )
\cr
D^{-1}_{11} &= \gamma {|\omega| \over q} + \left ( \chi_0 + \alpha^2 v(q)
\right ) q^2 \cr
&\approx \gamma {|\omega| \over q} + \chi q^{\eta} \ ,
}
\eqno{(12)}
$$
where $\chi = \chi_0 + \alpha^2 V_0$ for $\eta = 2$ and
$\chi = \alpha^2 V_0$ for $\eta \not= 2$.

Since the longitudinal part of the gauge field is screened, the transverse
part of the gauge field dominates the physics.
The one-loop self energy correction due to the transverse part of the
gauge field is calculated as (Fig.2) [1,12,20]
$$
\eqalign{
\Sigma ({\bf k}, i \omega) &= \int {d^2 q \over (2 \pi)^2} \ {d\nu \over 2 \pi}
\
\left | {{\bf k} \times {\hat {\bf q}} \over m} \right |^2
G_0 ({\bf k} + {\bf q}, i \omega +  i \nu) \ D_{11} ({\bf q}, i \nu) \cr
&\approx - i \ \lambda \ |\omega|^{2 \over 1 + \eta} \ {\rm sgn} (\omega) \ ,
}
\eqno{(13)}
$$
where
$$
\lambda = {v_F \over 4 \pi \ {\rm sin} ({2 \pi \over 1 + \eta}) \
\gamma^{\eta - 1 \over \eta + 1} \ \chi^{2 \over 1 + \eta} } \ ,
$$
and $G^{-1}_0 ({\bf k}, i \omega) = i \omega - \xi_{\bf k}$ ($\xi_{\bf k} =
{k^2 \over 2 m} - \mu$).
The self energy as a function of real frequency $\Omega$ (in the Minkowski
space)
can be obtained from the analytic continuation of
$\Sigma ({\bf k}, i \omega)$, {\it i.e.},
$\Sigma ({\bf k}, \Omega) =
\Sigma ({\bf k}, i \omega \rightarrow \Omega + i \delta)$.
Note that $|{\rm Im} \ \Sigma ({\bf k}, \Omega)| \propto
|\Omega|^{2 \over 1 + \eta} \gg |\Omega|$
for sufficiently small
$\Omega$ or $|\Omega| \ll \lambda^{\eta + 1 \over \eta - 1}$.
Therefore, the quasi-particle (the dressed fermion) is not well defined.

This can be also seen from the spectral function of fermions. The spectral
function can be obtained from the imaginary part of the retarded Green's
function: $A ({\bf k}, \Omega) = - {1 \over \pi} \ {\rm Im} \
G_R ({\bf k}, \Omega) = - {1 \over \pi} \ {\rm Im} \
G ({\bf k}, i \omega \rightarrow \Omega + i \delta)$, where
$G^{-1} ({\bf k}, i \omega) = G^{-1}_0 ({\bf k}, i \omega) -
\Sigma ({\bf k}, i \omega)$.
In the low frequency limit,
$$
A ({\bf k}, \Omega) \approx {1 \over \pi} \ {\lambda_2 \ |\Omega|^{2 \over 1 +
\eta}
\ {\rm sgn} (\Omega) \over (\lambda_1 \ |\Omega|^{2 \over 1 + \eta} - \xi_{\bf
k})^2
+ (\lambda_2 \ |\Omega|^{2 \over 1 + \eta})^2} \ ,
\eqno{(14)}
$$
where $\lambda_1 = \lambda \ {\rm cos} \left [ {\pi \over 2}
\left ( {\eta - 1 \over \eta + 1} \right ) \right ]$ and
$\lambda_2 = \lambda \ {\rm sin} \left [ {\pi \over 2}
\left ( {\eta - 1 \over \eta + 1} \right ) \right ]$.
Note that the maximum of $A ({\bf k}, \Omega)$ appears at
$\Omega \sim \left ( {\xi_{\bf k} \over \lambda_1} \right )^{1 + \eta \over
2}$.
However, the width of the broad peak is also order
$\Delta \Omega \sim
\left ( {\xi_{\bf k} \over \lambda_1} \right )^{1 + \eta \over 2}$.
Therefore, the Landau criterion for the existence of quasi-particles
($\Delta \Omega \ll \Omega$) is marginally violated.

If we assumed that there is a well-defined Fermi wave vector
$k_F = ({4 \pi n_f})^{1/2}$ and tried to fit the result to the usual
quasi-particle picture, the energy spectrum of the quasi-particle would be [1]
$$
\epsilon_{\bf k} \ \propto \ |k - k_F|^{1 + \eta \over 2}
\eqno{(15)}
$$
for $k$ sufficiently close to $k_F$.
{}From ${k_F \over m^*} = \left. {\partial \epsilon_{\bf k} \over \partial k}
\right |_{k = k_F}$, the effective mass diverges as
$$
m^* \ \propto \ |k - k_F|^{- {\eta - 1 \over 2}} \ \propto \
|\epsilon_{\bf k}|^{-{\eta - 1 \over \eta + 1}} \ .
\eqno{(16)}
$$
This suggests that at least some modifications to the conventional
Fermi-liquid theory are necessary as far as the one-particle Green's function
is concerned.

There have been also some non-perturbative calculations of the one-particle
Green's function [21-24], which were motivated by the singular perturbative
correction at low energies.
The results look very different from that obtained by the lowest order
perturbative calculation and even exponentially decaying one-particle
Green's function is found in the so-called eikonal limit [23].

{}From these results, one may doubt the validity of the quasi-particle
picture although a modified Fermi liquid description is proposed for the case
of the HFLL [1].
However, one should also remember that the one-particle Green's function
is not gauge invariant.
This can be easily seen in the path integral representation of the
one-particle Green's function [12,21] of a
fermion interacting with a gauge field,
{\it i.e.}, each path acquires a phase factor $e^{i \int^{t}_{0} \ dt'
{\bf a}({\bf r}, t') \cdot d {\bf r} / d t'}$ which is manifestly not
gauge invariant.
Therefore, it is very important to examine gauge-invariant quantities.
As the first example, we will calculate the polarization function for
$q \rightarrow 0$ case in the next section.

\vskip 0.5cm

\centerline{\bf III. THE TRANSVERSE POLARIZATION FUNCTION FOR $q \rightarrow
0$}
\centerline{\bf AND OPTICAL CONDUCTIVITY}

Let us consider a large $N$ generalized model of Eq.(8), where
$N$ is the number of species of fermions.
In this model, each fermion bubble carries a factor of $N$ and
each gauge field line gives a factor of $1/N$.
Thus, for example, $\Pi^0_{00}$ and $\Pi^0_{11}$ obtained in the previous
section should be multiplied by $N$.

In this section, we consider only the $q \rightarrow 0$ case of the transverse
polarization function: $\Pi_{11} ({\bf q} \rightarrow 0, i \nu)$.
However, the relevant diagrams are the same even for $q \not= 0$ case.
The leading order contribution is $\Pi^0_{11}$
which is proportional to $N$.
The relevant diagrams in the next order ({\it i.e.} $(1/N)^0$th order)
are given by Fig. 3 (a)-(g).
For convenience let us define the following quantities:
$\Pi^{(1)}_{11} = ({\rm a}) + ({\rm b})$ and
$\Pi^{(2)}_{11} = ({\rm c}) + ({\rm d})$.
The formal expressions of these quantities for $q \rightarrow 0$ case are
given by
$$
\Pi^{(1)}_{11} = - \int {d^2 k \over (2 \pi)^2} \
{d\omega \over 2 \pi} \ \left [ {k^2 - ({\bf k} \cdot {\hat {\bf q}})^2
\over m^2} \right ]
\ \Sigma ({\bf k}, i \omega) \
[ G_0 ({\bf k}, i \omega) ]^2 \ G_0 ({\bf k}, i \omega + i \nu) \ ,
\eqno{(17)}
$$
and
$$
\Pi^{(2)}_{11} = - \int {d^2 k \over (2 \pi)^2} \
{d\omega \over 2 \pi} \ \left [ {k^2 - ({\bf k} \cdot {\hat {\bf q}})^2
\over m^2} \right ]
\ \Sigma ({\bf k}, i \omega + i \nu) \
[ G_0 ({\bf k}, i \omega + i \nu) ]^2 \ G_0 ({\bf k}, i \omega) \ .
\eqno{(18)}
$$
These two equations can be rewritten as
$$
\eqalign{
\Pi^{(1)}_{11} &= - \int {d^2 k \over (2 \pi)^2} \
{d\omega \over 2 \pi} \ \left [ {k^2 - ({\bf k} \cdot {\hat {\bf q}})^2
\over m^2} \right ]
\ {\Sigma ({\bf k}, i \omega) \over i \nu} \cr
&\times
\left ( [ G_0 ({\bf k}, i \omega) ]^2 - G_0 ({\bf k}, i \omega)
\ G_0 ({\bf k}, i \omega + i \nu) \right ) \ ,
}
\eqno{(19{\rm a})}
$$
$$
\eqalign{
\Pi^{(2)}_{11} &= \int {d^2 k \over (2 \pi)^2} \
{d\omega \over 2 \pi} \ \left [ {k^2 - ({\bf k} \cdot {\hat {\bf q}})^2
\over m^2} \right ]
\ {\Sigma ({\bf k}, i \omega + i \nu) \over i \nu} \cr
&\times \left ( [ G_0 ({\bf k}, i \omega + i \nu) ]^2  -
G_0 ({\bf k}, i \omega + i \nu) \ G_0 ({\bf k}, i \omega) \right ) \ .
}
\eqno{(19{\rm b})}
$$
If we add (19a) and (19b), the first terms in each
polarization bubble are cancelled by each other and the remaning parts
give us
$$
\eqalign{
\Pi^{(1)}_{11} + \Pi^{(2)}_{11}
&= \int {d^2 k \over (2 \pi)^2} \
{d\omega \over 2 \pi} \ \left [ {k^2 - ({\bf k} \cdot {\hat {\bf q}})^2
\over m^2} \right ] \cr
&\times
{\Sigma ({\bf k}, i \omega) - \Sigma ({\bf k}, i \omega + i \nu) \over i \nu} \
G_0 ({\bf k}, i \omega) \ G_0 ({\bf k}, i \omega + i \nu) \ .
}
\eqno{(20)}
$$
{}From the above expression, it can be easily seen that
the contributions from (b) and (d) are automatically cancelled
because the self energy corrections in these diagrams are just the
same constants.

Next we consider the diagram given in Fig.3 (e).
Here we have to include the vertex correction for $q \rightarrow 0$ case
(Fig.4):
$$
\eqalign{
\Gamma_1 ({\bf k}, {\bf q} \rightarrow 0; i \omega, i\nu) &=
\int {d^2 q' \over (2 \pi)^2} \ {d\nu' \over 2 \pi} \
\left ( - {k_1 + q'_1 \over m} \right ) \
\left [ {k^2 - ({\bf k} \cdot {\hat {\bf q'}})^2
\over m^2} \right ] \cr
&\times
G_0 ({\bf k} + {\bf q'}, i \omega + i \nu') \
G_0 ({\bf k} + {\bf q'}, i \omega + i \nu' + i \nu) \
D_{11} ({\bf q'}, i \nu') \ .
}
\eqno{(21)}
$$
Then $\Pi^{(3)}_{11} ({\bf q} \rightarrow 0, i \nu)$ can be written as
$$
\eqalign{
\Pi^{(3)}_{11} &= - \int {d^2 k \over (2 \pi)^2} \ {d\omega \over 2 \pi} \
\left [ - {k_1 \over m} \right ] \
\Gamma_1 ({\bf k}, {\bf q} \rightarrow 0; i \omega, i\nu) \
G_0 ({\bf k}, i \omega) \ G_0 ({\bf k}, i \omega + i \nu) \cr
&= \Pi^{(3,1)}_{11} + \Pi^{(3,2)}_{11} \ ,
}
\eqno{(22)}
$$
where
$$
\eqalign{
\Pi^{(3,1)}_{11} &= - \int {d^2 k \over (2 \pi)^2} \ {d\omega \over 2 \pi} \
\left [ {k^2 - ({\bf k} \cdot {\hat {\bf q}})^2
\over m^2} \right ] \
G_0 ({\bf k}, i \omega) \ G_0 ({\bf k}, i \omega + i \nu) \cr
&\times
\int {d^2 q' \over (2 \pi)^2} \ {d\nu' \over 2 \pi} \
\left [ {k^2 - ({\bf k} \cdot {\hat {\bf q'}})^2
\over m^2} \right ] \cr
&\times
G_0 ({\bf k} + {\bf q'}, i \omega + i \nu') \
G_0 ({\bf k} + {\bf q'}, i \omega + i \nu' + i \nu) \
D_{11} ({\bf q'}, i \nu') \ ,
}
\eqno{(23)}
$$
and
$$
\eqalign{
\Pi^{(3,2)}_{11} &= - \int {d^2 k \over (2 \pi)^2} \ {d\omega \over 2 \pi} \
G_0 ({\bf k}, i \omega) \ G_0 ({\bf k}, i \omega + i \nu) \cr
&\times
\int {d^2 q' \over (2 \pi)^2} \ {d\nu' \over 2 \pi} \
\left ( q'_1 k_1 \over m^2 \right ) \
\left [ {k^2 - ({\bf k} \cdot {\hat {\bf q'}})^2
\over m^2} \right ] \cr
&\times G_0 ({\bf k} + {\bf q'}, i \omega + i \nu') \
G_0 ({\bf k} + {\bf q'}, i \omega + i \nu' + i \nu) \
D_{11} ({\bf q'}, i \nu') \ .
}
\eqno{(24)}
$$
Here we would like to point out that $\Pi^{(3,1)}_{11}$ is more singular
than $\Pi^{(3,2)}_{11}$. This can be easily seen from the fact that
$\Pi^{(3,2)}_{11}$ can be obtained by replacing
$k^2_1 / m^2 = \left [ {k^2 - ({\bf k} \cdot {\hat {\bf q}})^2
\over m^2} \right ]$ in the integrand of Eq.(23) by $q'_1 k_1 / m^2$.
Using $q'_1 = q'_{\parallel} \ {\rm sin} \ \theta_{\bf k q} +
q'_{\perp} \ {\rm cos} \ \theta_{\bf k q}$ and $\xi_{\bf k + q} \approx
\xi_{\bf k} + v_F q_{\parallel} + q^2_{\perp} / 2m$, one can do the
integrals over $q'_{\parallel}$ and $q'_{\perp}$ in Eq.(24).
Since the contribution from $q'_{\perp} \ {\rm cos} \ \theta_{\bf k q}$
term becomes an odd function of $q'_{\perp}$, this term vanishes.
By a formal manipulation, one can replace $q'_{\parallel}$ by
${q'}_{\perp}^2 / k_F$ so that $q'_1$ factor becomes effectively
$({q'}_{\perp}^2 / k_F) \ {\rm sin} \ \theta_{\bf kq}$.
Since the integrand is dominated by
$|\nu| \sim \left ( {\chi / \gamma} \right )
|q_{\perp}|^{1 + \eta}$ scaling given by the
pole of the gauge field propagator, replacing $k_1$ by $q'_1$ gives rise to an
additional factor which is proportional to $|\nu|^{2 \over 1 + \eta}$.
Therefore, $\Pi^{(3,2)}_{11}$ should be less singular than
$\Pi^{(3,1)}_{11}$ by the factor $|\nu|^{2 \over 1 + \eta}$ in the low
frequency
limit.

Note that $\Pi^{(3,1)}_{11}$ can be rewritten as
$$
\Pi^{(3,1)}_{11} = - \int {d^2 k \over (2 \pi)^2} \ {d\omega \over 2 \pi} \
\left [ {k^2 - ({\bf k} \cdot {\hat {\bf q}})^2
\over m^2} \right ] \
\Gamma_0 ({\bf k}, {\bf q} \rightarrow 0; i \omega, i\nu)
G_0 ({\bf k}, i \omega) \ G_0 ({\bf k}, i \omega + i \nu) \ ,
\eqno{(25)}
$$
where $\Gamma_0$ is the scalar vertex:
$$
\eqalign{
\Gamma_0 ({\bf k}, {\bf q}; i \omega, i\nu) &=
\int {d^2 q' \over (2 \pi)^2} \ {d\nu' \over 2 \pi} \
\left [ {k^2 - ({\bf k} \cdot {\hat {\bf q'}})^2
\over m^2} \right ] \cr
&\times
G_0 ({\bf k} + {\bf q'}, i \omega + i \nu') \
G_0 ({\bf k} + {\bf q'} + {\bf q}, i \omega + i \nu' + i \nu) \
D_{11} ({\bf q'}, i \nu') \ .
}
\eqno{(26)}
$$
{}From the relation,
$$
\eqalign{
\Sigma ({\bf k}, i \omega) - \Sigma ({\bf k}, i \omega + i \nu)
&= \int {d^2 q' \over (2 \pi)^2} \ {d\nu' \over 2 \pi} \
\left [ {k^2 - ({\bf k} \cdot {\hat {\bf q'}})^2
\over m^2} \right ] \cr
&\times
\left [ G_0 ({\bf k} + {\bf q'}, i \omega + i \nu') -
G_0 ({\bf k} + {\bf q'}, i \omega + i \nu' + i \nu) \right ]
\ D_{11} ({\bf q'}, i \nu') \cr
&= \int {d^2 q' \over (2 \pi)^2} \ {d\nu' \over 2 \pi} \
\left [ {k^2 - ({\bf k} \cdot {\hat {\bf q'}})^2
\over m^2} \right ] \ i \nu \cr
&\times G_0 ({\bf k} + {\bf q'}, i \omega + i \nu') \
G_0 ({\bf k} + {\bf q'}, i \omega + i \nu' + i \nu) \
D_{11} ({\bf q'}, i \nu') \ .
}
\eqno{(27)}
$$
we get the following identity:
$$
{\Sigma ({\bf k}, i \omega) - \Sigma ({\bf k}, i \omega + i \nu) \over
i \nu} = \Gamma_0 ({\bf k}, {\bf q} \rightarrow 0; i \omega, i\nu) \ .
\eqno{(28)}
$$
This is nothing but the Ward identity.
{}From Eqs.(20), (25), and (28), we have
$$
\Pi^{(1)}_{11} + \Pi^{(2)}_{11} + \Pi^{(3,1)}_{11} = 0 \ .
\eqno{(29)}
$$
Now the remaining piece is just $\Pi^{(3,2)}_{11}$.
Following the procedures of integration mentioned above, in the low frequency
limit,
we get
$$
\Pi^{(3,2)}_{11} \approx - {1 + \eta \over 4 \pi^2 \ (5 + \eta) \ {\rm sin}
\left (
{3 - \eta \over 1 + \eta} \pi \right ) } \ {v_F \over m} \
{\gamma^{3 - \eta \over 1 + \eta} \over \chi^{4 \over 1 + \eta} } \
|\nu|^{3 - \eta \over 1 + \eta} \ .
\eqno{(30)}
$$

Here it is worthwhile to compare this result with
$\Pi^{(1)}_{11} + \Pi^{(2)}_{11}$ and $\Pi^{(3,1)}_{11}$, {\it i.e.}, the
results before cancellation.
By a straightforward calculation, one can get
$$
\Pi^{(1)}_{11} + \Pi^{(2)}_{11}
\approx - {2 \ (1 + \eta) \over \pi \ (3 + \eta)} \ m \ v^2_F \ \lambda \
|\nu|^{ -{\eta - 1 \over \eta + 1} } \ .
\eqno{(31)}
$$
In order to calculate $\Pi^{(3,1)}_{11}$, the vertex correction should be
calculated.
The vertex correction $\Gamma_0 ({\bf k}, {\bf q} \rightarrow 0; i \omega,
i\nu)$
is found to be
$$
\Gamma_0 \approx
- {v_F \over \gamma} \ {1 \over 2 \pi \ {\rm sin} \left ( {2 \pi \over
1 + \eta} \right ) } \ {1 \over \nu} \
\left [ \ \left ( {|\omega| \gamma \over \chi} \right )^{2 \over 1 + \eta}
\ {\rm sgn} (\omega)
- \left ( {|\omega + \nu| \gamma \over \chi} \right )^{2 \over 1 + \eta}
\ {\rm sgn} (\omega + \nu) \ \right ] \ .
\eqno{(32)}
$$
Using Eqs.(25) and (32), $\Pi^{(3,1)}_{11}$ can be calculated as
$$
\Pi^{(3,1)}_{11} \approx {m \ v^3_F \over 2 \pi^2 \ {\rm sin} \left (
{2 \pi \over 1 + \eta} \right ) } \ \left ( {1 + \eta \over 3 + \eta} \right )
\ {1 \over \gamma^{\eta - 1 \over \eta + 1} \ \chi^{2 \over 1 + \eta} } \
|\nu|^{ -{\eta - 1 \over \eta + 1} } \ .
\eqno{(33)}
$$
Note that, as mentioned above, $\Pi^{(1)}_{11} + \Pi^{(2)}_{11}$ and
$\Pi^{(3,1)}_{11}$
are more singular than $\Pi^{(3,2)}_{11}$ by $|\nu|^{- {2 \over 1 + \eta}}$
in the low frequency limit.
The important point is that these singular terms are cancelled by each other
due to the Ward identity.

Now let us look at the diagrams of (f) and (g). Let $\Pi^{(4)}_{11} = ({\rm
f})$
and $\Pi^{(5)}_{11} = ({\rm g})$.
The formal expressions of these diagrams for $q \rightarrow 0$ case are given
by
$$
\eqalign{
\Pi^{(4)}_{11} &=
\int {d^2 q' \over (2 \pi)^2} \ {d\nu' \over 2 \pi} \
{d^2 k' \over (2 \pi)^2} \ {d\omega' \over 2 \pi} \
{d^2 k'' \over (2 \pi)^2} \ {d\omega'' \over 2 \pi} \cr
&\times
\left [ {{\bf k'} \cdot {\bf k''} - ({\bf k'} \cdot {\hat {\bf q'}}) \
({\bf k''} \cdot {\hat {\bf q'}}) \over m^2} \right ]^2 \
D_{11} ({\bf q'}, i \nu') \ D_{11} ({\bf q'}, i \nu' + i \nu) \cr
&\times
G_0 ({\bf k'}, i \omega') \ G_0 ({\bf k'}, i \omega' + i \nu) \
G_0 ({\bf k'} - {\bf q'}, i \omega' - i \nu') \cr
&\times
G_0 ({\bf k''}, i \omega'') \ G_0 ({\bf k''}, i \omega'' + i \nu) \
G_0 ({\bf k''} - {\bf q'}, i \omega'' - i \nu') \ ,
}
\eqno{(34)}
$$
and
$$
\eqalign{
\Pi^{(5)}_{11} &=
\int {d^2 q' \over (2 \pi)^2} \ {d\nu' \over 2 \pi} \
{d^2 k' \over (2 \pi)^2} \ {d\omega' \over 2 \pi} \
{d^2 k'' \over (2 \pi)^2} \ {d\omega'' \over 2 \pi} \cr
&\times
\left [ {{\bf k'} \cdot {\bf k''} - ({\bf k'} \cdot {\hat {\bf q'}}) \
({\bf k''} \cdot {\hat {\bf q'}}) \over m^2} \right ]^2 \
D_{11} ({\bf q'}, i \nu') \ D_{11} ({\bf q'}, i \nu' + i \nu) \cr
&\times
G_0 ({\bf k'}, i \omega') \ G_0 ({\bf k'}, i \omega' + i \nu) \
G_0 ({\bf k'} - {\bf q'}, i \omega' - i \nu') \cr
&\times
G_0 ({\bf k''}, i \omega'') \ G_0 ({\bf k''}, i \omega'' + i \nu) \
G_0 ({\bf k''} + {\bf q'}, i \omega'' +  i \nu' + i \nu) \ .
}
\eqno{(35)}
$$
By changing variables as
${\bf q'} \rightarrow - {\bf q'}$, $\nu' \rightarrow - \nu' - \nu$ and
using $D_{11} (- {\bf q'}, - i \nu') = D_{11} ({\bf q'}, i \nu')$, we get
$$
\eqalign{
\Pi^{(4)}_{11} + \Pi^{(5)}_{11} &= {1 \over 2}
\int {d^2 q' \over (2 \pi)^2} \ {d\nu' \over 2 \pi} \
D_{11} ({\bf q'}, i \nu') \ D_{11} ({\bf q'}, i \nu' + i \nu) \cr
&\times
\Biggl [
{d^2 k \over (2 \pi)^2} \ {d\omega \over 2 \pi} \ {k_1 \over m} \
\left ( {k \ {\rm sin} \ \theta_{{\bf k} {\bf q'}} \over m} \right )^2
G_0 ({\bf k}, i \omega) \ G_0 ({\bf k}, i \omega + i \nu) \cr
&\times
\left ( G_0 ({\bf k} + {\bf q'}, i \omega + i \nu' + i \nu) +
G_0 ({\bf k} - {\bf q'}, i \omega - i \nu') \right ) \Biggr ]^2 \ ,
}
\eqno{(36)}
$$
where $\theta_{{\bf k} {\bf q'}}$ is the angle between ${\bf k}$ and ${\bf
q'}$.
In the low frequency limit, we get
$$
\Pi^{(4)}_{11} + \Pi^{(5)}_{11} \approx
- c_1 \ {v_F \over m} \ {\gamma^{3 - \eta \over 1 + \eta}
\over \chi^{4 \over 1 + \eta} } \ |\nu|^{3 - \eta \over 1 + \eta} \ ,
\eqno{(37)}
$$
where $c_1$ is a constant.
One can also show that
$$
\eqalign{
\Pi^{(4)}_{11} &\approx - c_0 \ {m \ v^3_F \over \gamma^{\eta - 1 \over \eta +
1} \
\chi^{2 \over 1 + \eta} } \
|\nu|^{ -{\eta - 1 \over \eta + 1} } \ , \cr
\Pi^{(5)}_{11} &\approx c_0 \ {m \ v^3_F \over \gamma^{\eta - 1 \over \eta + 1}
\
\chi^{2 \over 1 + \eta} } \
|\nu|^{ -{\eta - 1 \over \eta + 1} } - c_1 \ {v_F \over m} \
{\gamma^{3 - \eta \over 1 + \eta}
\over \chi^{4 \over 1 + \eta} } \ |\nu|^{3 - \eta \over 1 + \eta} \ ,
}
\eqno{(38)}
$$
where $c_0$ is a constant.
That is, there is also a cancellation between the singular parts of
$\Pi^{(4)}_{11}$ and $\Pi^{(5)}_{11}$.

Gathering all the previous informations and using
$\Pi^0_{11} ({\bf q} \rightarrow 0, i \nu) = {N n \over m}$, we can
conclude that
$$
\Pi_{11} \ \approx \ {N n \over m} -
c_2 \ {k_F \over m^2} \ {\gamma^{3 - \eta \over 1 + \eta}
\over \chi^{4 \over 1 + \eta} } \ |\nu|^{3 - \eta \over 1 + \eta}
\eqno{(39)}
$$
up to $(1/N)^0$th order, where $c_2$ is a constant.

In order to calculate the optical conductivity, we have to consider
the bubble diagrams with two external lines that represent the coupling
to the external vector potential $A_{\mu}$ while the internal gauge field
lines are due to $a_{\mu}$. There are additional diagrams generated by
$\psi^{\dagger} a_{\mu} A^{\mu} \psi$ vertex.
All the additional diagrams except one (shown in Fig.5 (a)) vanish
due to the symmetry of the integrand. A typical diagram which vanishes is
shown in Fig.5 (b).
It turns out that the diagram represented by Fig.5 (a) gives an imaginary
part which is higher order in frequency compared to
$|\nu|^{3 - \eta \over 1 + \eta}$ so that it is irrelevant in the low
freqency limit.
Now we can use the imaginary part of the transverse polarization function
in the Minkowski space $\Pi_{11} ({\bf q} \rightarrow 0, \Omega) =
\Pi_{11} ({\bf q} \rightarrow 0, i \nu
\rightarrow \Omega + i \delta)$ to calculate the real part of the optical
conductivity:
$$
{\rm Re} \ \sigma (\Omega) =
- e^2 \ {{\rm Im} \ \Pi_{11} (\Omega) \over \Omega} \ .
\eqno{(40)}
$$
{}From Eq.(39), ${\rm Re} \ \sigma (\Omega)$ is given by
$$
{\rm Re} \ \sigma (\Omega) \ \propto \
{e^2 k_F \over m^2} \ {\gamma^{3 - \eta \over 1 + \eta}
\over \chi^{4 \over 1 + \eta} } \
\Omega^{- 2 \left ( {\eta - 1 \over \eta + 1 } \right ) } \ .
\eqno{(41)}
$$
If there were no cancellation, the result would look quite different.
For example, if we did not consider the vertex correction, the result
from $\Pi^{(1)}_{11} + \Pi^{(2)}_{11}$ would be
$$
{\rm Re} \ \sigma_{nv} (\Omega) \ \propto \
{e^2 m v^3_F \over \gamma^{\eta - 1 \over \eta + 1} \
\chi^{2 \over 1 + \eta} } \
\Omega^{- {2 \eta \over 1 + \eta}} \ ,
\eqno{(42)}
$$
where $\sigma_{nv}$ represents the conductivity without vertex correction.

Now we are going to show that the right answer given by Eq.(41) is
consistent with a modified Drude formula if we assume that
the transport scattering rate (which is the inverse of the transport time
$\tau_{\rm tr}$) of the fermion is given by $\Gamma_{\rm tr} (\Omega)
\propto {1 \over N} \ {1 \over m k_F} \ ( \gamma^{3 - \eta \over 1 + \eta} /
\chi^{4 \over 1 + \eta} ) \
\Omega^{4 \over 1 + \eta}$.

First of all, for later convenience, let us calculate the inverse of the
transport time $\tau^0_{\rm tr}$ of the fermion [12] using
the imaginary part of the self energy $\Sigma ({\bf k}, \Omega)$.
For this purpose, we can just include the factor $1 - {\rm cos} \ \Theta =
2 \ {\rm sin}^2 (\Theta / 2)$ in the integrand of the expression for
${\rm Im} \ \Sigma ({\bf k}, \Omega)$, where $\Theta$ is the angle between
the wave vector of the fermion and that of the gauge field [12].
Using the fact that ${\rm sin} \ (\Theta / 2)
\approx q / 2 k_F$ and
$q \sim \left ( {\gamma \Omega \over \chi} \right )^{1 \over 1 + \eta}$
inside the integral [12], we get
$$
{1 \over \tau^0_{\rm tr}} \ \propto \
{1 \over N} \ {1 \over m k_F} \ {\gamma^{3 - \eta \over 1 + \eta} \over
\chi^{4 \over 1 + \eta} } \
\Omega^{4 \over 1 + \eta}
\eqno{(43)}
$$
Therefore, we will essentially show that our result of the optical
conductivity is consistent with the identification of
$\Gamma_{\rm tr} = 1 / \tau^0_{\rm tr}$ or
$\tau_{\rm tr} = \tau^0_{\rm tr}$ in a modified Drude formula.

The Drude formula that is appropriate to the large $N$ generalized model
is given by
$$
{\rm Re} \ \sigma (\Omega) = {N n e^2 \over m} \ {\Gamma_{\rm tr} \over
\Omega^2 + \Gamma^2_{\rm tr}} \ .
\eqno{(44)}
$$
In the large $N$ limit, if we assume
$\Gamma_{\rm tr} = 1 / \tau^0_{\rm tr} \propto 1/N$,
$$
{\rm Re} \ \sigma (\Omega) \approx {N n e^2 \over m} \ {\Gamma_{\rm tr} \over
\Omega^2} \ \propto \ {e^2 v_F \over m} \ {\gamma^{3 - \eta \over 1 + \eta}
\over
\chi^{4 \over 1 + \eta}} \
\Omega^{- 2 \left ( {\eta - 1 \over \eta + 1 } \right ) } \ .
\eqno{(45)}
$$
This is the same result as that of Eq.(41).
The result of Eq.(42) can be reproduced in the same way if we assume that
$\Gamma_{\rm tr} (\Omega)
\propto {1 \over N} \ (m v^3_F) ( \gamma^{-{\eta - 1 \over \eta + 1}}
\ \chi^{-{2 \over 1 + \eta}} )
\ \Omega^{2 \over 1 + \eta}$ which is essentially
the imaginary part of the self energy $\Sigma ({\bf k}, \Omega)$.
Therefore, the optical conductivity is consistent with the choice of
$1 / \tau^0_{\rm tr}$ rather than just the
naive scattering rate (given by the self energy) as the transport scattering
rate.
Since the singular contribution, which gives Eq.(42), is
cancelled by the vertex correction, we can again say that the leading singular
behaviors of one-particle properties do not show up in the optical
conductivity.

For finite temperature, one can replace $\Omega$ by $T$ in $\Gamma_{\rm tr}$.
Note that the DC-limit of the optical conductivity
${\rm Re} \ \sigma (\Omega \rightarrow 0) = {N n e^2 \over m}
{1 \over \Gamma_{\rm tr}}$
cannot be obtained by the $1/N$ expansion.
However, one can infer the DC-limit by assuming that the full
${\rm Re} \ \sigma (\Omega)$ is given by Eq.(44) (with
$\Gamma_{\rm tr} = \Gamma_{\rm tr} (T)$) which is consistent with the
result of the large-$N$ limit of the optical conductivity.
If $\Gamma_{\rm tr} \propto T^{4 \over 1 + \eta}$ was used, one would get
${\rm Re} \ \sigma (T) \propto T^{- {4 \over 1 + \eta}}$ [12].
One the other hand, one would get
${\rm Re} \ \sigma_{nv} (T) \propto T^{- {2 \over 1 + \eta}}$ if
$\Gamma_{\rm tr} \propto T^{2 \over 1 + \eta}$ was used.
In Ref.[19], the authors concluded that the resistivity of the
system is proportional to $T^{2 / 3}$ for the short-range interaction
$(\eta = 2)$ and this is consistent with the latter case.
Therefore, our result is in disagreement with their conclusion about
the resistivity.

\vskip 0.5cm

\centerline{\bf IV. THE TRANSVERSE POLARIZATION FUNCTION FOR FINITE $q \ll
k_F$}
\centerline{\bf AND NON-RENORMALIZATION OF THE GAUGE FIELD PROPAGATOR}

It is not easy to find the polarization function for arbitrary
${\bf q}$ and $\nu$.
However, some simplifications can be made for $q \ll k_F$ case.
In this section, we calculate $\Pi_{11} ({\bf q}, i \nu)$ for
finite $q \ll k_F$ up to two-loop order. We set $N=1$ first, and discuss
the extension to the large-$N$ case later.

First of all, $\Pi^{(1)}_{11}$ and $\Pi^{(2)}_{11}$ for finite ${\bf q}$
have the following form:
$$
\eqalign{
\Pi^{(1)}_{11} &= - \int {d^2 k \over (2 \pi)^2} \
{d\omega \over 2 \pi} \ \left [ {k^2 - ({\bf k} \cdot {\hat {\bf q}})^2
\over m^2} \right ]
\ \Sigma ({\bf k}, i \omega) \
[ G_0 ({\bf k}, i \omega) ]^2 \ G_0 ({\bf k} + {\bf q}, i \omega + i \nu)
\ , \cr
\Pi^{(2)}_{11} &= - \int {d^2 k \over (2 \pi)^2} \
{d\omega \over 2 \pi} \ \left [ {k^2 - ({\bf k} \cdot {\hat {\bf q}})^2
\over m^2} \right ]
\ \Sigma ({\bf k} + {\bf q}, i \omega + i \nu) \cr
&\times
[ G_0 ({\bf k} + {\bf q}, i \omega + i \nu) ]^2 \ G_0 ({\bf k}, i \omega) \ .
}
\eqno{(46)}
$$
Using the similar method as that used in section III, one can obtain
$$
\eqalign{
\Pi^{(1)}_{11} + \Pi^{(2)}_{11}
&\approx \int {d^2 k \over (2 \pi)^2} \
{d\omega \over 2 \pi} \ \left [ {k^2 - ({\bf k} \cdot {\hat {\bf q}})^2
\over m^2} \right ] \
G_0 ({\bf k}, i \omega) \ G_0 ({\bf k} + {\bf q}, i \omega + i \nu) \cr
&\times
{\Sigma ({\bf k}, i \omega) - \Sigma ({\bf k} +
{\bf q}, i \omega + i \nu) \over i \nu - v_F q \ {\rm cos} \
\theta_{{\bf k} {\bf q}} } \ .
}
\eqno{(47)}
$$
Next we should consider the vertex correction (Fig.4) for finite ${\bf q}$:
$$
\eqalign{
\Gamma_1 ({\bf k}, {\bf q}; i \omega, i\nu) &=
\int {d^2 q' \over (2 \pi)^2} \ {d\nu' \over 2 \pi} \
A ({\bf k}, {\bf q}, {\bf q'}) \
B ({\bf k}, {\bf q}, {\bf q'}) \cr
&\times
G_0 ({\bf k} + {\bf q'}, i \omega + i \nu') \
G_0 ({\bf k} + {\bf q'} + {\bf q}, i \omega + i \nu' + i \nu) \
D_{11} ({\bf q'}, i \nu') \ ,
}
\eqno{(48)}
$$
where
$$
\eqalign{
A &= - {k_1 + q'_1 + q_1 / 2 \over m} = - {k_1 + q'_1 \over m} \cr
B &= {1 \over m} \left [ ({\bf k} + {\bf q'}/2) \cdot ({\bf k} + {\bf q} + {\bf
q'}/2) -
({\bf k} + {\bf q'}/2) \cdot {\hat {\bf q'}} \
({\bf k} + {\bf q} + {\bf q'}/2) \cdot {\hat {\bf q'}} \right ] \ .
}
\eqno{(49)}
$$
For $q \ll k_F$ and $|{\bf k}| \approx k_F$, the following approximation
can be made
$$
B \approx {k^2 - ({\bf k} \cdot {\hat {\bf q'}})^2 \over m} \ .
\eqno{(50)}
$$
Using this approximation, one can show that
$$
\eqalign{
\Pi^{(3)}_{11} &= - \int {d^2 k \over (2 \pi)^2} \ {d\omega \over 2 \pi} \
\left [ - {k_1 \over m} \right ] \
\Gamma_1 ({\bf k}, {\bf q}; i \omega, i\nu) \
G_0 ({\bf k}, i \omega) \ G_0 ({\bf k} + {\bf q}, i \omega + i \nu) \cr
&\approx \Pi^{(3,3)}_{11} + \Pi^{(3,4)}_{11} \ ,
}
\eqno{(51)}
$$
where
$$
\eqalign{
\Pi^{(3,3)}_{11} &= - \int {d^2 k \over (2 \pi)^2} \ {d\omega \over 2 \pi} \
\left [ {k^2 - ({\bf k} \cdot {\hat {\bf q}})^2
\over m^2} \right ] \
\Gamma_0 ({\bf k}, {\bf q}; i \omega, i\nu) \
G_0 ({\bf k}, i \omega) \ G_0 ({\bf k} + {\bf q}, i \omega + i \nu)
\ , \cr
\Pi^{(3,4)}_{11} &= - \int {d^2 k \over (2 \pi)^2} \ {d\omega \over 2 \pi} \
G_0 ({\bf k}, i \omega) \ G_0 ({\bf k} + {\bf q}, i \omega + i \nu) \cr
&\times
\int {d^2 q' \over (2 \pi)^2} \ {d\nu' \over 2 \pi} \
\left ( q'_1 k_1 \over m^2 \right ) \
\left [ {k^2 - ({\bf k} \cdot {\hat {\bf q'}})^2
\over m^2} \right ] \cr
&\times G_0 ({\bf k} + {\bf q'}, i \omega + i \nu') \
G_0 ({\bf k} + {\bf q'} + {\bf q}, i \omega + i \nu' + i \nu) \
D_{11} ({\bf q'}, i \nu') \ .
}
\eqno{(52)}
$$

First, let us calculate the scalar vertex part
$\Gamma_0 ({\bf k}, {\bf q}; i \omega, i\nu)$.
We use $\xi_{{\bf k} + {\bf q'}} \approx \xi_{\bf k} + v_F q'_{\parallel} +
{q'}^2_{\perp} / 2m$ and
$\xi_{{\bf k} + {\bf q'} + {\bf q}} \approx \xi_{\bf k} + v_F q'_{\parallel}
+ v_F q \ {\rm cos} \ \theta_{\bf k q} + {q q'_{\perp} \over m} \
{\rm sin} \ \theta_{\bf k q} + {q'}^2_{\perp} / 2m$
(where $q'_{\parallel} = q' \
{\rm cos} \ \theta_{{\bf k} {\bf q'}}$ and
$q'_{\perp} = q' \ {\rm sin} \ \theta_{{\bf k} {\bf q'}}$) to perform the
integral in Eq.(26).
Using the fact that the important region of $q'$ is the order of
$\nu^{1 \over 1 + \eta} \ll 1$ so that $q' / k \approx q' / k_F \ll 1$,
we conclude [23,27,28] that
$q'_{\parallel} / k_F \approx ( q'_{\perp} / k_F )^2$ and
we can approximate the
gauge field propagator as
$D_{11} ({\bf q'}, i\nu') \approx
1 / ( \gamma |\nu'|/|q'_{\perp}| + \chi |q'_{\perp}|^{\eta} )$.
After performing $q'_{\parallel}$ integral, we get
$$
\eqalign{
\Gamma_0 ({\bf k}, {\bf q}; i \omega, i\nu)
&\approx
- i v_F \int {d\nu' \over 2 \pi} \ \int {d q'_{\perp} \over 2 \pi} \
( {\rm sgn} (\omega + \nu') - {\rm sgn} (\omega + \nu + \nu') ) \cr
&\times
{1 \over i \nu - v_F q \ {\rm cos} \ \theta_{\bf k q} -
{q q'_{\perp} \over m} \ {\rm sin} \ \theta_{\bf k q} } \
{1 \over \gamma {|\nu'| \over |q'_{\perp}|} + \chi |q'_{\perp}|^{\eta}} \ .
}
\eqno{(53)}
$$
Now $\nu'$ integral gives
$$
\eqalign{
\Gamma_0 ({\bf k}, {\bf q}; i\omega, i\nu)
&\approx
- {v_F \over \gamma} {1 \over \pi^2} \int^{k_F}_{-k_F} d q'_{\perp} \
{|q'_{\perp}| \over \nu + i v_F q \ {\rm cos} \ \theta_{\bf k q} +
i {q q'_{\perp} \over m} \ {\rm sin} \ \theta_{\bf k q} } \cr
&\times
\left [ {\rm ln} \left
( 1 + {|\omega| \gamma \over |q'_{\perp}|^{1+\eta} \chi} \right )
{\rm sgn} (\omega) - {\rm ln} \left
( 1 + {|\omega + \nu| \gamma \over |q'_{\perp}|^{1+\eta} \chi} \right )
{\rm sgn} (\omega + \nu) \right ] \ .
}
\eqno{(54)}
$$
By changing variables, one can get the following formula.
$$
\eqalign{
&\hskip 0.5cm \Gamma_0 ({\bf k}, {\bf q}; i \omega, i\nu) \cr
&\approx
- {v_F \over \gamma} \ {1 \over \pi^2} \
{1 \over \nu + i v_F q \ {\rm cos} \ \theta_{{\bf k} {\bf q}} } \cr
&\times \Biggl [
\left ( {|\omega| \gamma \over \chi} \right )^{2 \over 1 + \eta} \
F \left ( \omega, \ {(q/m) \ {\rm sin} \ \theta_{{\bf k} {\bf q}} \over
v_F q \ {\rm cos} \ \theta_{{\bf k} {\bf q}} - i \nu} \
\left [ {|\omega| \gamma \over \chi} \right ]^{1 \over 1 + \eta}
\right ) \ {\rm sgn} (\omega) \cr
&- \left ( {|\omega + \nu| \gamma \over \chi} \right )^{2 \over 1 + \eta} \
F \left ( \omega + \nu, \ {(q/m) \ {\rm sin} \ \theta_{{\bf k} {\bf q}} \over
v_F q \ {\rm cos} \ \theta_{{\bf k} {\bf q}} - i \nu} \
\left [ {|\omega + \nu| \gamma \over \chi} \right ]^{1 \over 1 + \eta}
\right ) \ {\rm sgn} (\omega + \nu) \Biggr ] \ .
}
\eqno{(55)}
$$
Here $F (\omega, x)$ is defined as
$$
F (\omega, x) = \int^{y_c}_{-y_c} dy \ |y| \
{{\rm ln} \ (1 + |y|^{-1-\eta}) \over 1 + x y} \ ,
\eqno{(56)}
$$
where $y_c = k_F \left ( {\chi \over |\omega| \gamma}
\right )^{1 \over 1 + \eta}$.
It can be easily shown that $q \rightarrow 0$ limit of Eq.(55) is
given by Eq.(32).
On the other hand, the self energy can be rewritten as
$$
\Sigma ({\bf k}, \omega) \approx - i {v_F \over \pi^2 \gamma} \
\left ( |\omega| \gamma \over \chi \right )^{2 \over 1 + \eta}
{\rm sgn} (\omega) \
F(\omega, 0)
\eqno{(57)}
$$
Collecting these results, it can be shown that
$$
\eqalign{
\Pi^{(1)}_{11} + \Pi^{(2)}_{11} + \Pi^{(3,3)}_{11}
&\approx - \int {d^2 k \over (2 \pi)^2} \ {d\omega \over 2 \pi} \
\left [ {k^2 - ({\bf k} \cdot {\hat {\bf q}})^2
\over m^2} \right ] \
G_0 ({\bf k}, i \omega) \ G_0 ({\bf k} + {\bf q}, i \omega + i \nu) \cr
&\times {i v_F \over \pi^2 \gamma} \ {1 \over v_F q \ {\rm cos} \
\theta_{{\bf k} {\bf q}} - i \nu} \
\biggl [ I (\omega) - I (\omega + \nu) \biggr ] \ ,
}
\eqno{(58)}
$$
where
$$
\eqalign{
I (\omega) &= \left ( |\omega| \gamma \over \chi \right )^{2 \over 1 + \eta}
{\rm sgn} (\omega) \cr
&\times
\left [ F \left ( \omega, \ {(q/m) \ {\rm sin} \ \theta_{{\bf k} {\bf q}} \over
v_F q \ {\rm cos} \ \theta_{{\bf k} {\bf q}} - i \nu} \
\left [ {|\omega| \gamma \over \chi} \right ]^{1 \over 1 + \eta}
\right ) - F (\omega, 0) \right ] \ .
}
\eqno{(59)}
$$

The integrals in Eq.(58) can be evaluated as the following.
Using $\int d^2 k / (2 \pi)^2 =
(m / 2 \pi) \int d \xi_{\bf k} \ \int d \theta_{\bf k q} / 2 \pi$,
one can perform $\xi_{\bf k}$ integral easily.
The angular integral over $\theta_{\bf k q}$ can be done by contour
integration, which requires long algebraic manipulations.
The remaining $\omega$ integral and the $y$ integral in $I (\omega)$ of
Eq.(59) can be evaluated by scaling the integration variables and
expanding the integrand in some limits.
More details of the calculation will be demonstrated in the later
evalution of the density-density correlation function (see the
discussions about Eqs.(68)-(70) in section V) which can be more
easily calculated.
First, for $|\nu| \ll v_F q$,
$$
\Pi^{(1)}_{11} + \Pi^{(2)}_{11} + \Pi^{(3,3)}_{11} \approx c_3 \
{m^2 v^3_F \over \gamma} \ {|\nu| \over v_F q} \
{\left ( \gamma |\nu| / \chi \right )^{4 \over 1 + \eta}
\over k^3_F q} \ ,
\eqno{(60)}
$$
while, in the other limit $|\nu| \gg v_F q$, we get
$$
\Pi^{(1)}_{11} + \Pi^{(2)}_{11} + \Pi^{(3,3)}_{11} \approx c_4 \
{m^2 v^3_F \over \gamma} \ {q v_F \over |\nu|} \ {q \over k_F} \
\left [ {\left ( \gamma / \chi \right )^{2 \over 1 + \eta} \over
m |\nu|^{\eta - 1 \over \eta + 1} } \right ]^2 \ ,
\eqno{(61)}
$$
where $c_3$ and $c_4$ are dimensionless constants.

The calculation of $\Pi^{(3,4)}_{11}$ can be also done by the similar
method used in the evaluation of $\Pi^{(3,3)}_{11}$.
First, for $|\nu| \ll v_F q$, we get
$$
\Pi^{(3,4)}_{11} \approx
- {m^2 v^3_F \over \gamma} \ {|\nu| \over v_F q} \ \left [
c_5 \ {\left ( \gamma |\nu| / \chi \right )^{2 \over 1 + \eta}
\over k^2_F}
+ c_6 \ {\left ( \gamma |\nu| / \chi \right )^{3 \over 1 + \eta}
\over k^2_F q} \right ] \ ,
\eqno{(62)}
$$
whereas, in the other limit $|\nu| \gg v_F q$,
$$
\Pi^{(3,4)}_{11} \approx
- {1 + \eta \over 4 \pi^2 (5 + \eta)} \
{1 \over {\rm sin} \left ( {4 \pi \over 1 + \eta} \right ) } \
{v_F \over m} \
{\gamma^{3 - \eta \over 1 + \eta} \over \chi^{4 \over 1 + \eta}} \
|\nu|^{3 - \eta \over 1 + \eta}
- c_7 \ {m^2 v^3_F \over \gamma} \ {v_F q^2
\over m^2 (\chi / \gamma)^{3 \over 1 + \eta} |\nu|^{3 \over 1 + \eta}} \ ,
\eqno{(63)}
$$
where $c_5, c_6$ and $c_7$ are dimensionless constants.

{}From the above results, it can be shown that
$|\Pi^{(1)}_{11} + \Pi^{(2)}_{11} + \Pi^{(3,3)}_{11}| < |\Pi^{(3,4)}_{11}|$
for relevant limits.
Therefore, the imaginary part of the transverse polarization function
$\Pi_{11} ({\bf q}, \Omega)$ (in the Minkowski space)
up to two-loop order is given by the following formulae.
For $\Omega \ll v_F q$, we get
$$
{\rm Im} \ \Pi_{11} ({\bf q}, \Omega) \approx
- {m v^2_F \over 2 \pi} \ {\Omega \over v_F q} \
\left [ 1 - a \ {m v_F \over \gamma} \
{\left ( \gamma \Omega / \chi \right )^{2 \over 1 + \eta} \over k^2_F}
- b \ {m v_F \over \gamma} \
{\left ( \gamma \Omega / \chi \right )^{3 \over 1 + \eta} \over k^2_F q}
\right ] \ ,
\eqno{(64)}
$$
where $a$ and $b$ are dimensionless constants.
Note that the correction is small as far as $1 < \eta \le 2$ is concerned.
On the other hand, for $\Omega \gg v_F q$, we have
$$
{\rm Im} \ \Pi_{11} ({\bf q}, \Omega) \approx
- {1 + \eta \over 8 \pi^2 (5 + \eta)} \
{1 \over {\rm sin} \left ( {2 \pi \over 1 + \eta} \right ) } \
{v_F \over m} \
{\gamma^{3 - \eta \over 1 + \eta} \over \chi^{4 \over 1 + \eta}} \
\Omega^{3 - \eta \over 1 + \eta} \
\left [ 1 + c \ m v^3_F \ \left ( {\chi \over \gamma} \right )^{1 \over 1 +
\eta}
{q^2 \over \Omega^{2 \eta + 3 \over \eta + 1}} \right ] \ ,
\eqno{(65)}
$$
where $c$ is a dimensionless constant.

For $\Omega > v_F q$, there is no contribution to
${\rm Im} \ \Pi_{11}$ from the free fermion bubble because
the regime is outside the particle-hole continuum.
Therefore, any non-zero contribution to ${\rm Im} \ \Pi_{11}$
for $\Omega \gg v_F q$ entirely comes from the gauge field correction.
Note that the first term in Eq.(65) dominates for $\Omega >
(m v^3_F)^{1 + \eta \over 2 \eta + 3} (\chi / \gamma)^{1 \over 2 \eta + 3}
q^{2 \eta + 2 \over 2 \eta + 3}$.
On the other hand, the second term becomes more important for $v_F q \ll \Omega
< (m v^3_F)^{1 + \eta \over 2 \eta + 3} (\chi / \gamma)^{1 \over 2 \eta + 3}
q^{2 \eta + 2 \over 2 \eta + 3}$ so that
${\rm Im} \ \Pi_{11} \propto v^4_F {\gamma^{2 - \eta \over 1 + \eta} \over
\chi^{3 \over 1 + \eta}} {q^2 \over \Omega^{3 \eta \over 1 + \eta}}$
in this regime.
As we approach the line given by $\Omega = v_F q$,
${\rm Im} \ \Pi_{11}$ becomes $v_F^{4 + \eta \over 1 + \eta}
{\gamma^{2 - \eta \over 1 + \eta} \over \chi^{3 \over 1 + \eta}}
q^{2 - \eta \over 1 + \eta}$ as a function of q.

In the case of $\Omega \ll v_F q$, the free fermion bubble gives
${\rm Im} \ \Pi^0_{11} = - {m v^2_F \over 2 \pi} {\Omega \over v_F q}$.
Note that
${\rm Im} \ \Pi_{11} ({\bf q}, \Omega) \approx
- {m v^2_F \over 2 \pi} \ {\Omega \over v_F q} \
\left [ 1 - a \ {m v_F \over \gamma} \
{\left ( \gamma \Omega / \chi \right )^{2 \over 1 + \eta} \over k^2_F}
\right ]$ for $\Omega < (\chi / \gamma) q^{1 + \eta}$
and
${\rm Im} \ \Pi_{11} \approx
- {m v^2_F \over 2 \pi} \ {\Omega \over v_F q} \
\left [ 1 - b \ {m v_F \over \gamma} \
{\left ( \gamma \Omega / \chi \right )^{3 \over 1 + \eta} \over k^2_F q}
\right ]$ for $(\chi / \gamma) q^{1 + \eta} < \Omega \ll v_F q$.
It is gratifying to note that, along the line $\Omega = v_F q$, the correction
to ${\rm Im} \ \Pi_{11}$ given by the above expression agrees in its $q$
dependence
with that obtained by approaching from $\Omega \gg v_F q$ given in the last
paragraph.
In any case, the corrections are small compared to the free fermion result
for $1 < \eta \le 2$.

Using the result of $\Pi_{11}$ for $|\nu| \ll v_F q$,
we can discuss the issue of
the renormalization of the gauge field propagator.
Recall that the dispersion relation of the gauge field obtained from the
one-loop
correction is given by $|\nu| \sim (\chi / \gamma) q^{1 + \eta}$ [1,12,13],
which is below the line of $|\nu| = v_F q$ for sufficiently small $q$.
Along the line of $|\nu| \sim (\chi / \gamma) q^{1 + \eta}$, one can easily
see that the correction to $\Pi^0_{11}$ is smaller by ${m v_F \over \gamma}
\left ( {q \over k_F} \right )^2$.
Therefore, the gauge field propagator is not renormalized up to two-loop
order.
As mentioned in the introduction, non-renormalization of the gauge field
propagator was first discussed by Polchinski within a self-consistent
argument and without vertex correction.
In Ref.[19], the authors discussed the relevance of
$\Gamma^{(3)} (a_{\mu})$ and $\Gamma^{(4)} (a_{\mu})$, which are coefficients
of the $a^3$ and $a^4$ terms in the expansion of the effective action of the
gauge field. They concluded that $\Gamma^{(3)} (a_{\mu})$ and
$\Gamma^{(4)} (a_{\mu})$ are irrelevant so that the gauge field is not
renormalized. Since the two-loop diagrams we considered are generated from
$\Gamma^{(4)} (a_{\mu})$, our calculation is consistent with their conclusion.
By analogy, we expect that $\Pi^{(4)}_{11}$ and $\Pi^{(5)}_{11}$ are irrelevant
for the renormalization of the gauge field because these are generated from
$\Gamma^{(3)} (a_{\mu})$.
We also directly evaluated $\Gamma^{(3)} (a_{\mu})$ and confirmed the argument
of Ref.[19].
Therefore, one can expect that the gauge field is not renormalized up to
$(1/N)^0$th order in the $1/N$ expansion.
That is, the RPA calculation gives the leading contributions in the low energy
limit.

\vskip 0.5cm

\centerline{\bf V. THE DENSITY-DENSITY CORRELATION FUNCTION FOR FINITE
$q \ll k_F$}

The polarization function for the density channel $\Pi_{00} ({\bf q}, \Omega)$
can be also calculated in a similar way as used in section IV.
In this section, we consider the two-loop corrections given by Fig.3
(a)-(e) and finite $q \ll k_F$ case.
The sum of the contributions from the self-energy corrections given by
Fig.3 (a)-(d) can be written as
$$
\Pi^{(1)}_{00}
\approx \int {d^2 k \over (2 \pi)^2} \
{d\omega \over 2 \pi} \
G_0 ({\bf k}, i \omega) \ G_0 ({\bf k} + {\bf q}, i \omega + i \nu) \
{\Sigma ({\bf k}, i \omega) - \Sigma ({\bf k} +
{\bf q}, i \omega + i \nu) \over i \nu - v_F q \ {\rm cos} \
\theta_{{\bf k} {\bf q}} } \ ,
\eqno{(66)}
$$
while the contribution given by Fig.3 (e), which comes from the vertex
correction, can be also written as
$$
\Pi^{(2)}_{00} = - \int {d^2 k \over (2 \pi)^2} \ {d\omega \over 2 \pi} \
\Gamma_0 ({\bf k}, {\bf q}; i \omega, i\nu) \
G_0 ({\bf k}, i \omega) \ G_0 ({\bf k} + {\bf q}, i \omega + i \nu) \ .
\eqno{(67)}
$$
Using Eqs.(55) and (57), it can be shown that
$$
\eqalign{
\Pi^{(1)}_{00} + \Pi^{(2)}_{00}
&\approx - \int {d^2 k \over (2 \pi)^2} \ {d\omega \over 2 \pi} \
G_0 ({\bf k}, i \omega) \ G_0 ({\bf k} + {\bf q}, i \omega + i \nu) \cr
&\times {i v_F \over \pi^2 \gamma} \ {1 \over v_F q \ {\rm cos} \
\theta_{{\bf k} {\bf q}} - i \nu} \
\biggl [ I (\omega) - I (\omega + \nu) \biggr ] \ ,
}
\eqno{(68)}
$$
where $I (\omega)$ is given by Eq.(59).
Using $\int d^2 k / (2 \pi)^2 =
(m / 2 \pi) \int d \xi_{\bf k} \ \int d \theta_{\bf k q} / 2 \pi$,
one can easily perform $\xi_{\bf k}$ integral, which generates the
additional factor $v_F q \ {\rm cos} \ \theta_{{\bf k} {\bf q}} - i \nu$
in the denomenator of the integrand of Eq.(68).
Recalling that $I (\omega)$ also has an angle dependence
$\theta_{{\bf k} {\bf q}}$, one can perform
the angular integral over $\theta_{\bf k q}$ by contour
integration, which requires long algebraic manipulations.
After rescaling the $\omega$ integral by a new variable $x$ and the
$y$ integral in $I (\omega)$ (see Eqs.(56) and (59)) by newly defined
$y$, we get
$$
\eqalign{
\Pi^{(1)}_{00} + \Pi^{(2)}_{00}
&\approx {2 k^3_F \over \pi^3 \gamma} {|\nu| \over v^2_F q^2} \
\int^1_0 \ dx \ \int^1_0 \ dy \ y \ {\rm ln} \ \left ( 1 +
{x \beta^{1 + \eta} \over y^{1 + \eta}} \right ) \cr
&\times
\left [ \ {|\alpha| \over (1 + \alpha^2) \sqrt{1 + \alpha^2 + y^2}} -
{|\alpha| \over (1 + \alpha^2)^{3/2}} \ \right ] \ ,
}
\eqno{(69)}
$$
where $\alpha = {\nu \over v_F q}$ and $\beta = {1 \over k_F}
\left ( {|\nu| \gamma \over \chi} \right )^{1 \over 1 + \eta}$.
%For $1 + \alpha^2 \gg \beta^2$, only $y \ll \sqrt{1 + \alpha^2}$ is
%important
%and we may expand the first term in the square bracket.
In the small frequency $\nu$ limit, the parameter integrals can be done,
yielding
$$
\eqalign{
\Pi^{(1)}_{00} + \Pi^{(2)}_{00} &\approx
- {a_1 \over k^{\eta - 2}_F \chi} \
{|\alpha|^3 \over (1 + \alpha^2)^{3/2}} \cr
&- {1 + \eta \over 4 \pi^2 (5 + \eta)} {1 \over {\rm sin}
\left ( {4 \pi \over 1 + \eta} \right )} {1 \over k_F \gamma}
{1 \over v_F q}
\left ( {\gamma |\nu| \over \chi} \right )^{4 \over 1 + \eta}
{\alpha^2 \over (1 + \alpha^2)^{5/2}} \ ,}
\eqno{(70)}
$$
where $a_1$ is an undetermined constant.
This formula is valid for all ratios of $q$ and $\nu$, as long as both
are small.
Note that the first term gives only an analytic contribution, which
also arises in the usual Fermi liquid theory.
Similar methods can be used to produce a somewhat more complicated
formula valid for all $\alpha$ for the transverse polarization function
$\Pi_{11}$ (for example, Eqs.(52) and (58) can be evaluated by a similar
method).

After dropping the analytic contribution, we combine the
free fermion contribution and perform analytic continuation to
get, for $\Omega \ll v_F q$,
$$
{\rm Im} \ \Pi_{00} (q, \Omega) \approx - {m \over 2 \pi}
\ {\Omega \over v_F q} \ \left [ \ 1 -
{1 + \eta \over 4 \pi (5 + \eta)} {1 \over {\rm cos}
\left ( {\eta - 1 \over \eta + 1} \pi \right )}
{1 \over k_F m}
{\gamma^{3 - \eta \over 1 + \eta} \over \chi^{4 \over 1 + \eta}}
\Omega^{3 - \eta \over 1 + \eta}
\left ( {\Omega \over v_F q} \right )^2
\ \right ] \ ,
\eqno{(71)}
$$
and for $\Omega \gg v_F q$,
$$
{\rm Im} \ \Pi_{00} (q, \Omega) \approx
- {1 + \eta \over 8 \pi^2 (5 + \eta)} {1 \over {\rm sin}
\left ( {2 \pi \over 1 + \eta} \right )} {1 \over k_F}
{\gamma^{3 - \eta \over 1 + \eta} \over \chi^{4 \over 1 + \eta}}
\Omega^{3 - \eta \over 1 + \eta}
\left ( {v_F q \over \Omega} \right )^2 \ .
\eqno{(72)}
$$
Note that ${\rm Im} \ \Pi_{11} (q \rightarrow 0, \Omega) =
{\Omega^2 \over v^2_F q^2}
{\rm Im} \ \Pi_{00} (q \rightarrow 0, \Omega)$
is satisfied.
Therefore, both of ${\rm Im} \ \Pi_{11} (q \rightarrow 0, \Omega)$ and
${\rm Im} \ \Pi_{00} (q \rightarrow 0, \Omega)$ give the same answer
for the optical conductivity given by Eq.(41).

\vskip 0.5cm

\centerline{\bf VI. COMPARISION TO THE FERMI LIQUID THEORY}

In section III, it was shown that the resulting conductivity is
consistent with a modefied Drude formula.
In this section, we try to fit this result to
the Fermi liquid theory framework to extract informations
about the Fermi liquid parameters and examine whether the gauge field
induces some singular or divergent parameters.
In the Fermi liquid theory, the conductivity for $N$ species of fermions is
given by [31]
$$
\sigma (\Omega) = {N n e^2 \over m^*} \
{\tau \over 1 - i \Omega \tau (m / m^*)} \ ,
\eqno{(73)}
$$
or
$$
{\rm Re} \ \sigma (\Omega) = {N n e^2 \over m} \
{\Gamma_{\rm tr} \over \Omega^2 + \Gamma^2_{\rm tr}} \ ,
\eqno{(74)}
$$
where $\Gamma_{\rm tr} = \Gamma_{\rm sc} {m^* \over m}$,
$\Gamma_{\rm sc} = 1 / \tau$ is the scattering rate and $\tau$ is the
scattering time.
Here $m^*$ is the effective mass of the fermion.
Using the fact $\Gamma_{\rm tr} \propto 1/N$ in the large $N$ limit, we get
$$
{\rm Re} \ \sigma (\Omega) \approx {N n e^2 \over m} \
{\Gamma_{\rm tr} \over \Omega^2} \ .
\eqno{(75)}
$$
Comparing the above result with Eq.(41) which is a result of the $1/N$
expansion, we can again identify $\Gamma_{\rm tr}$ with $1 / \tau^0_{\rm tr}$
given in Eq.(43).
Therefore, we can conclude that
$\Gamma_{\rm tr} = \Gamma_{\rm sc} {m^* \over m}$
scales as $\Omega^{4 \over 1 + \eta}$ after including $1/N$ corrections due
to the gauge field fluctuations.

In the following we will directly compare our perturbative result for
$\Pi_{00}$ with the density-density correlation function in the Fermi
liquid theory.
Our goal is to find out whether the perturbative result
can be consistent with a Fermi
liquid theory made up of quasi-particles with a divergent effective
mass $m^*$ as suggested, for example, by Eq.(16).
First we consider the limit $\Omega = 0, q \rightarrow 0$, where it
is well known that the Fermi liquid theory predicts
$$
\Pi_{00} ({\bf q} \rightarrow 0, \Omega = 0) =
{\Pi^*_{00} ({\bf q} \rightarrow 0, \Omega = 0) \over
1 + f_{0s} \ \Pi^*_{00} ({\bf q} \rightarrow 0, \Omega = 0)} \ ,
\eqno{(76)}
$$
where
$\Pi^*_{00} = - \int {d^2 p \over (2 \pi)^2}
{n^0_{\bf p} - n^0_{\bf p - q} \over
\Omega - (\epsilon^*_{\bf p} - \epsilon^*_{\bf p - q})}$ is the free fermion
response fuunction with an effective mass $m^*$ and $f_{0s}$ is the
angular average of the Fermi liquid interaction parameter $f_{\bf pp'}$.
In two dimensions, for small q limit,
$$
\Pi^*_{00}({\bf q}, \Omega) =
- {m^* \over 2 \pi} \left ( 1 - {x \over \sqrt{x^2 - 1}} \ \theta (x^2 - 1)
+ i {x \over \sqrt{1 - x^2}} \ \theta (1 - x^2) \right ) \ ,
\eqno{(77)}
$$
where $x = \Omega / v^*_F q$.
In Euclidean space, the above formula can be reduced to
$$\Pi^*_{00} ({\bf q}, i\nu) = -
{m^* \over 2 \pi} \left ( 1 - {|\alpha| \over \sqrt{1 + {\alpha}^2}} \right ) \
,
\eqno{(78)}
$$
where $\alpha = \nu / v^*_F q$.
Since $\Pi^*_{00} ({\bf q} \rightarrow 0, \Omega = 0) \propto m^*$,
the fact that $\Pi_{00} ({\bf q} \rightarrow 0, \Omega = 0)$ is
not enhanced implies that $f_{0s}$ is a finite constant.
However, this does not imply that the leading order term in the
perturbative expansion of $f_{0s}$ is finite.
In fact, it is clear from an expansion of Eq.(76) that
if the leading order correction to $m$
is singular, then the contribution to $f_{0s}$ at the same order
should be also singular since $\Pi_{00}$ has no singular correction
in the lowest order perturbation theory.

Next we consider the full $q, \Omega$ dependence of $\Pi_{00}$ for
small $q$ and $\Omega$.
We are motivated by the belief that, in the Fermi liquid theory,
${\rm Im} \ \Pi_{00} ({\bf q}, \Omega)$ should exhibit the edge of the
particle-hole continuum along the line $\Omega = v^*_F q$.
However, when $\Omega \not= 0$, a simple formula such as
Eq.(76) does not exist for $\Pi_{00} ({\bf q}, \Omega)$.
In particular, $\Pi_{00} ({\bf q}, \Omega)$ in general depends on
the higher moment angular average of the Landau functions, and
not just $f_{0s}$.
Nevertheless, the Fermi liquid theory makes a precise prediction for
$\Pi_{00} ({\bf q}, \Omega)$ for all $q, \Omega$ in terms of $m^*$ and
the interaction parameter $f_{\bf pp'}$.
This is given by the quantum Boltzmann equation for the quasi-particle
distribution function $n_{\bf p} = n^0_{\bf p} + \delta n_{\bf p}$
in the Fermi liquid theory, where $n^0_{\bf p}$ is the
distribution function for the free fermion system with an
effective mass $m^*$:
$$
\eqalign{
&\left [ \ \Omega -
(\epsilon^*_{{\bf p} + {\bf q}/2} -
\epsilon^*_{{\bf p} - {\bf q}/2}) \ \right ]
\delta n_{\bf p} \cr
&- (n^0_{{\bf p} + {\bf q}/2} -
n^0_{{\bf p} - {\bf q}/2})
\left [ \ U ({\bf q}, \Omega)
+ \int {d^2 p' \over (2 \pi)^2} \
f_{\bf pp'} \ \delta n_{\bf p'} ({\bf q}, \Omega) \ \right ] = 0 \ . }
\eqno{(79)}
$$
Here $\epsilon^*_{\bf p}$ is the quasi-particle energy, $U ({\bf q}, \Omega)$
is the external potential, and $f_{\bf pp'}$ is the Fermi-liquid interaction
parameter.
The linear response of $\delta n_{\bf p}$ to the external potential
can be calculated from Eq.(79) (to the first order in $f_{\bf pp'}$):
$$
\eqalign{
\delta n_{\bf p} ({\bf q}, \Omega) &= \left [ \ c_{\bf p} +
\int {d^2 p' \over (2 \pi)^2} \ c_{\bf p} f_{\bf pp'} c_{\bf p'} \ \right ]
U ({\bf q}, \Omega) \cr
c_{\bf p} &= {n^0_{{\bf p} + {\bf q}/2} - n^0_{{\bf p} - {\bf q}/2} \over
\Omega - (\epsilon^*_{{\bf p} + {\bf q}/2} -
\epsilon^*_{{\bf p} - {\bf q}/2})} \ . }
\eqno{(80)}
$$
The change in the density of the fermions $\delta \rho ({\bf q}, \Omega)
= \int {d^2 p \over (2 \pi)^2} \ \delta n_{\bf p} ({\bf q}, \Omega)$
is given by
$$
\eqalign{
{\delta \rho ({\bf q}, \Omega) \over U ({\bf q}, \Omega)}
&= - \Pi_{00} ({\bf q}, \Omega) \cr
&= \int {d^2 p \over (2 \pi)^2} \
{n^0_{\bf p} - n^0_{\bf p - q} \over
\Omega - (\epsilon^*_{\bf p} - \epsilon^*_{\bf p - q})}
+ \int {d^2 p \ d^2 p' \over (2 \pi)^4} \ c_{\bf p} f_{\bf pp'} c_{\bf p'}
+ \cdots \ , }
\eqno{(81)}
$$
where $\cdots$ represents the higher order terms in $f_{\bf pp'}$.
The second term is just the diagram given in Fig.3 (e), but with a
frequency independent interaction $f_{\bf pp'}$.

Let us now examine what happens to the edge in the particle-hole
continuum according to our perturbative results.
The gauge interaction may induce non-zero Fermi-liquid interaction
function $f_{\bf pp'}$
and a change in the Fermi velocity $\delta v_F$.
{}From Eq.(78) and Eq.(81), a change in the Fermi velocity $\delta v_F$
and the appearance of the Fermi liquid interaction parameter
induce the following change in the density-density correlation function:
$$
\delta \Pi_{00} = - {\delta v_F \over v_F} \left ( - \Pi^*_{00}
+ {k_F \over 2\pi v_F} {|\alpha| \over (1 + {\alpha}^2)^{3/2}} \right )
- \int {d^2 p \ d^2 p' \over (2 \pi)^4} \ c_{\bf p} f_{\bf pp'}
c_{\bf p'} \ .
\eqno{(82)}
$$
If we assume a power law behavior for
$f_{\bf pp'} \sim {1 \over |{\bf p} - {\bf p'}|^{\lambda}}$ with
$\lambda < 1$ ({\it i.e.}, finite $f_{0s}$), one can show that
the second term in Eq.(82) cannot produce
the singular term $(1 + {\alpha}^2)^{-3/2}$ near ${\alpha}^2 = -1$.
To prove this argument, let us perform the integration over $|{\bf p}|$ and
$|{\bf p'}|$ in the small $q$ limit, yielding
$$
\int {d^2 p \ d^2 p' \over (2 \pi)^4} \ c_{\bf p} f_{\bf pp'}
c_{\bf p'} =
{4 k^2_F \over (2 \pi)^4} \int d \theta_{\bf p q} \ d \theta_{\bf p' q} \
{q^2 \ {\rm cos} \ \theta_{\bf p q} \ {\rm cos} \ \theta_{\bf p' q}
\ f_{\bf p p'}
\over (\Omega - v_F q \ {\rm cos} \ \theta_{\bf p q})
(\Omega - v_F q \ {\rm cos} \ \theta_{\bf p' q})} \ ,
\eqno{(83)}
$$
where $\theta_{\bf p q}$ ($\theta_{\bf p' q}$) is the angle between {\bf p}
and {\bf q} (${\bf p'}$ and ${\bf q}$).
In order to obtain the leading singularity near $\Omega = v_F q$,
the above expression can be further simplified:
$$
\eqalign{
&\int {d^2 p \ d^2 p' \over (2 \pi)^4} \ c_{\bf p} f_{\bf pp'}
c_{\bf p'} \cr
&= {4 k^2_F \over (2 \pi)^4 v^2_F}
\int d \theta_{\bf p q} \ d \theta_{\bf p' q} \
{f_{\bf p p'} \over \left [ \ \left ({\Omega \over v_F q} - 1 \right )
+ {1 \over 2} \theta^2_{\bf p q} \ \right ] \
\left [ \ \left ({\Omega \over v_F q} - 1 \right )
+ {1 \over 2} \theta^2_{\bf p' q} \ \right ]} \ .}
\eqno{(84)}
$$
For $f_{\bf p p'} \propto
{1 \over |\theta_{\bf p q} - \theta_{\bf p' q}|^{\lambda}}$ with $\lambda < 1$,
the above integral can be estimated through a scaling argument.
We find
$$
\int {d^2 p \ d^2 p' \over (2 \pi)^4} \ c_{\bf p} f_{\bf pp'}
c_{\bf p'} \propto
{1 \over \left ( {\Omega \over v_F q} - 1 \right )^{2 + \lambda \over 2}} \ ,
\eqno{(85)}
$$
which is less divergent than $(1 + \alpha^2)^{-3/2}$ term
that leads to $\left ( {\Omega \over v_F q} - 1 \right )^{-3/2}$ divergence.
Thus there is no cancellation between the first and the second terms in
Eq.(82).
If $\delta v_F$ diverges at small frequencies, we can conclude that
$\delta \Pi_{00}$ will
diverge in the limit $\nu \rightarrow 0$ with $\nu / v_F q$ fixed,
which contradicts to our two-loop result from Eq.(71) that shows no
such divergent term.
Similar results also hold for the transverse
current-current response function.

The argument above assumes a power law behavior for
$f_{\bf p p'} \propto
{1 \over |\theta_{\bf p q} - \theta_{\bf p' q}|^{\lambda}}$.
As $\lambda \rightarrow 1$, another possibility needs to be considered,
namely $f_{\bf {\hat p} {\hat p}'} \propto
\delta ({\bf \hat p} - {\bf \hat p'})$.
This satisfies the condition that $f_{0s}$ is finite.
{}From Eq.(84) it is clear that this will lead to a term of order
$(1 + \alpha^2)^{-3/2}$ which may cancel the first term in Eq.(82).
However, in this case, we shall argue that, at least at zero
temperature, $f_{\bf {\hat p} {\hat p}'}
= \zeta \ \delta ({\bf \hat p} - {\bf \hat p'})$
is equivalent to a shift in the Fermi velocity by $v_F \rightarrow
v_F + \zeta k_F / (2 \pi)^2$.
At zero temperature the excitation can be described by a distortion
of the Fermi surface in the direction ${\bf \hat p}$ by an
amount $\delta \nu_{\bf \hat p} = \int d |{\bf p}| \ \delta n_{\bf p}$.
The original Landau's expression of the free energy density takes the form:
$$
\eqalign{
\delta F &= \int {d^2 p \over (2 \pi)^2} \
v_F (|{\bf p}| - k_F) \delta n_{\bf p}
+ {1 \over 2} \int {d^2 p \ d^2 p' \over (2 \pi)^4} \
f_{\bf p p'} \delta n_{\bf p} \delta n_{\bf p'} \cr
&= \int {k_F \ d {\bf \hat p} \over (2 \pi)^2} \
{1 \over 2} v_F (\delta \nu_{\bf \hat p})^2
+ {1 \over 2} \int {k^2_F \ d {\bf \hat p} \ d {\bf {\hat p}'}
\over (2 \pi)^4} \
f_{\bf {\hat p} {\hat p}'} \delta \nu_{\bf \hat p}
\delta \nu_{\bf \hat p'} \ .}
\eqno{(86)}
$$
It is then clear that $f_{\bf {\hat p} {\hat p}'}
= \zeta \ \delta ({\bf \hat p} - {\bf \hat p'})$ is equivalent to
$v_F \rightarrow v_F + \zeta k_F / (2 \pi)^2$.
The same result can be also obtained by performing an
integral over $|{\bf p}|$
in Eq.(79), which leads to
$$
(\Omega - v_F q \ {\rm cos} \ \theta) \
\delta \nu_{\bf \hat p}
- q \ {\rm cos} \ \theta
\left [ \ U ({\bf q}, \Omega)
+ \int {k_F \ d {\bf {\hat p}'} \over (2 \pi)^2} \
f_{\bf {\hat p} {\hat p}'} \
\delta \nu_{\bf {\hat p}'} \ \right ] = 0
\eqno{(87)}
$$
in the small $q$ limit.
Thus we see that, at zero temperature, all response functions
to an external perturbation can be described by a Landau theory
with a non-divergent effective mass in the small $q$ limit.
However, it is also possible that the same response function can be
described by a Landau-Fermi-liquid theory of which both effective mass
and $f_{\bf p p'}$ have divergent perturbative corrections.

An examination of Eq.(70) shows that after analytic continuation, the factor
$(1 + \alpha^2)^{-5/2}$ diverges at $\Omega = v_F q$, even though its
coefficient
vanishes for $\Omega \rightarrow 0$.
In the following we attempt an interpretation of the result.
We can write our perturbative result Eq.(70) as, near $\Omega = v_F q$,
$$
{\rm Im} \ \Pi_{00} ({\bf q}, \Omega) =
{\rm Im} \ \Pi^0_{00} ({\bf q}, \Omega)
+ \alpha_0 {\partial \ {\rm Im} \ \Pi^0_{00} ({\bf q}, \Omega)
\over \partial \Omega}
+ \gamma_0 {\partial^2 \ {\rm Im} \ \Pi^0_{00} ({\bf q}, \Omega)
\over \partial \Omega^2} \ ,
\eqno{(88)}
$$
where $\Pi_{00}^0$ is given by Eq.(77) with $m^* \rightarrow m$, and
$$
\eqalign{
\alpha_0 &= {a_2 \over k^{\eta - 2}_F \chi} q \ , \cr
\gamma_0 &= {1 + \eta \over 8 \pi^2 (5 + \eta)} {1 \over {\rm cos}
\left ( {2 \pi \over 1 + \eta} \right )} {1 \over k_F \gamma}
{1 \over v_F q}
\left ( {\gamma \Omega \over \chi} \right )^{4 \over 1 + \eta}
q^2 \ ,}
\eqno{(89)}
$$
where $a_2$ is a constant.
The existence of
$\partial \ {\rm Im} \ \Pi^0_{00} ({\bf q}, \Omega)
/ \partial \Omega$ term in Eq.(88) signifies that there is
a finite non-singular (see $\alpha_0$ in Eq.(89)) shift in $v_F$, which
also arises in the usual Fermi liquid theory.
To interpret the second derivative term, we note that Eq.(88) is
consistent with (apart from the term proportional to $\alpha_0$)
$$
{\rm Im} \ \Pi_{00} ({\bf q}, \Omega) = {1 \over 2}
\left [ \ {\rm Im} \ \Pi^0_{00} ({\bf q}, \Omega + \Gamma) +
{\rm Im} \ \Pi^0_{00} ({\bf q}, \Omega - \Gamma) \ \right ]
\eqno{(90)}
$$
if $\Gamma = \sqrt{2 \gamma_0}$.
We recall that ${\rm Im} \ \Pi^0_{00} ({\bf q}, \Omega)$ has a discontinuity
at $\Omega = v_F q$, corresponding to the edge of the particle-hole
continuum.
Eq.(90) has the natural interpretation of a smearing of the discontinuity
at a shifted (due to a shift in $v_F$) edge of the particle-hole
continuum by the amount $\Gamma$.
Setting $v_F q \propto \Omega$, we find that
$$
\Gamma \ \propto \ \Omega^{1 + {3 - \eta \over 2 + 2 \eta}} \ .
\eqno{(91)}
$$
Note that for $\eta < 3$, $\Gamma < \Omega$ so that the above
picture is a self-consistent one.
We also note that $\Gamma$ is proportional to the square root of the
coupling constant or $1 / N$, and is therefore non-analytic.
We are not certain if any further physical meaning can be ascribed to
the energy scale $\Gamma$.

\vskip 0.5cm

\centerline{\bf VII. CONCLUSION}

In this paper we studied properties of gauge-invariant correlation
functions in a two-dimensional fermion system coupled to a gauge field.
We find the physical picture emerged from those gauge-invariant
correlation functions to be very different from those obtained from
gauge-dependent one-particle Green's function.
The corrections to the Fermi-liquid two-particle correlation functions
are found to be non-divergent and sub-leading to
the Fermi-liquid contributions
up to two-loop order, and there is no need to go beyond the
perturbation theory at this order.

However, it is still possible that singular corrections to
the gauge-invariant two-particle correlation functions may appear in
some special cases, such as $q = 2 k_F$.
Also, since we do not have quasi-particles to serve as the underpinning
of the Fermi-liquid-like behavior for $\Pi_{00}$ and $\Pi_{11}$, it
is possible that singularity shows up in some other response functions.
Nevertheless, the perturbative result should serve as a test for
any theory such as renormalization group analysis [26] which attempts
to go beyond perturbation theory.

Finally we would like to comment on the implication of our results
to the HTSC. Even though our results suggest that the two-particle
Green's functions of fermions are Fermi-liquid-like for small $q$ and
$\Omega$, it does not mean that the gauge field formulation of the
$t-J$ model (in relation to the normal state properties of HTSC)
leads to the Fermi-liquid
interpretation of the normal state of HTSC.
In the problem of the $t-J$ model, there are bosons as well as fermions
which are interacting with a gauge field [12].
In fact, the presence of fermions and bosons in this problem
came from the no-doulble-occupancy constraint on the electrons.
It has been also regarded as a way of describing the spin-charge
seperation induced by the strong correlation effects.
In the paper of Nagaosa and Lee [12], they clearly demonstrated
that the anomalous transport properties are due to the bosons.
That is, the presence of the bosons plays an important role in the
non-Fermi-liquid behaviors of the normal state of HTSC.
However, in this paper we considered only the fermions interacting
with a gauge field.

\vskip 0.5cm

\centerline{\bf ACKNOWLEDGMENTS}

We would like to thank B. I. Halperin for helpful discussions
and important comments on an early version of this manuscript.
We are also grateful to B. Altshuler, S. Chakravarty, A. Millis,
N. Nagaosa, and P. Stamp for discussions.
YBK and XGW are supported by NSF
grant No. DMR-9022933.
AF and PAL are supported by NSF grant No. DMR-9216007.

\vfill\vfill\vfill
\break

\vskip 0.5cm

\centerline{\bigbf References}

\vskip 0.5cm

\item{*} On leave from Department of Applied Physics, University of Tokyo,
Hongo, Bunkyo-ku, Tokyo 113, Japan.
\item{[1]} B. I. Halperin, P. A. Lee, and N. Read,
Phys. Rev. {\bf B} {\bf 47}, 7312 (1993).
\item{[2]} V. Kalmeyer and S. C. Zhang, Phys. Rev. {\bf B} {\bf 46},
9889 (1992).
\item{[3]} S. H. Simon and B. I. Halperin, Phys. Rev. {\bf B} {\bf 48},
17368 (1993);
S. H. Simon and B. I. Halperin, Phys. Rev. {\bf B} {\bf 50},
1807 (1994);
Song He, S. H. Simon, and B. I. Halperin, Phys. Rev. {\bf B} {\bf 50},
1823 (1994).
\item{[4]} J. K. Jain, Phys. Rev. Lett. {\bf 63}, 199 (1989);
Phys. Rev. {\bf B} {\bf 41}, 7653 (1990); Adv. Phys. {\bf 41}, 105 (1992).
\item{[5]} A. Lopez and E. Fradkin, Phys. Rev. {\bf B} {\bf 44},
5246 (1991); Phys. Rev. Lett. {\bf 69}, 2126 (1992).
\item{[6]} R. L. Willet, M. A. Paalanen, R. R. Ruel, K. W. West,
L. N. Pfeiffer, and D. J. Bishop, Phys. Rev. Lett. {\bf 65}, 112 (1990).
\item{[7]} R. L. Willet, R. R. Ruel, M. A. Paalanen, K. W. West,
and L. N. Pfeiffer, Phys. Rev. {\bf B} {\bf 47}, 7344 (1993).
\item{[8]} R. R. Du, H. L. Stormer, D. C. Tsui, L. N. Pfeiffer,
and K. W. West, Phys. Rev. Lett. {\bf 70}, 2944 (1993).
\item{[9]} W. Kang, H. L. Stormer, L. N. Pfeiffer, K. W. Baldwin,
and K. W. West, Phys. Rev. Lett. {\bf 71}, 3850 (1993).
\item{[10]} D. R. Leadley, R. J. Nicholas, C. T. Foxon, and J. J. Harries,
Phys. Rev. Lett. {\bf 72}, 1906 (1994).
\item{[11]} V. J. Goldman, B. Su, and J. K. Jain, Phys. Rev. Lett. {\bf 72},
2065 (1994).
\item{[12]} N. Nagaosa and P. A. Lee, Phys. Rev. Lett. {\bf 64}, 2550 (1990);
P. A. Lee and N. Nagaosa, Phys. Rev. {\bf B} {\bf 46}, 5621 (1992).
\item{[13]} L. B. Ioffe and A. I. Larkin,
Phys. Rev. {\bf B} {\bf 39}, 8988 (1989).
\item{[14]} P. A. Lee, Phys. Rev. Lett. {\bf 63}, 680 (1989).
\item{[15]} L. B. Ioffe and P. B. Wiegmann,
Phys. Rev. Lett. {\bf 65}, 653 (1990).
\item{[16]} L. B. Ioffe and G. Kotliar,
Phys. Rev. {\bf B} {\bf 42}, 10348 (1990).
\item{[17]} T. Holstein, R. E. Norton, and P. Pincus,
Phys. Rev. {\bf B} {\bf 8}, 2649 (1973).
\item{[18]} M. Yu Reizer, Phys. Rev. {\bf B} {\bf 39}, 1602 (1989);
Phys. Rev. {\bf B} {\bf 40}, 11571 (1989).
\item{[19]} Junwu Gan and Eugene Wong, Phys. Rev. Lett. {\bf 71},
4226 (1993).
\item{[20]} B. Blok and H. Monien, Phys. Rev. {\bf B} {\bf 47}, 3454 (1993).
\item{[21]} B. L. Altshuler and L. B. Ioffe, Phys. Rev. Lett. {\bf 69},
2979 (1992).
\item{[22]} D. V. Khveshchenko, R. Hlubina, and T. M. Rice,
Phys. Rev. {\bf B} {\bf 48}, 10766 (1993).
\item{[23]} D. V. Khveshchenko and P. C. E. Stamp, Phys. Rev. Lett. {\bf 71},
2118 (1993); Phys. Rev. {\bf B} {\bf 49}, 5227 (1994).
\item{[24]} H. -J. Kwon, A. Houghton, and J. B. Marston,
Phys. Rev. Lett. {\bf 73}, 284 (1994).
\item{[25]} Y. B. Kim and X.-G. Wen, Phys. Rev. {\bf B} {\bf 50}, XXXX (1994).
\item{[26]} C. Nayak and F. Wilczek, Nucl. Phys. {\bf B 417}, 359 (1994).
\item{[27]} L. B. Ioffe, D. Lidsky, and B. L. Altshuler,
Phys. Rev. Lett. {\bf 73}, 472 (1994).
\item{[28]} J. Polchinski, Nucl. Phys. {\bf B 422}, 617 (1994).
\item{[29]} We would like to remark that $\delta \ {\rm Im} \ \Pi^{s}_{11}$
does not modify the RPA dressed gauge field propagator even though this
gauge-dependent correction violates the Fermi-liquid criterion.
This is due to the fact that $\delta \ {\rm Im} \Pi^{s}_{11}$ becomes
less important than the free fermion result ${\rm Im} \ \Pi^{0}_{11}$
along the line $\Omega \propto q^3$ (which corresponds to the dispersion
relation of the gauge field) for small $q$ and $\Omega$ [19,23,28].
\item{[30]} L. B. Ioffe and V. Kalmeyer, Phys. Rev. {\bf B} {\bf 44}, 750
(1991).
\item{[31]} D. Pines and P. Nozieres, {\it The theory of quantum liquids},
{\it volume 1} (Benjamin, Reading, Massachusetts, 1966).

\vfill\vfill\vfill
\break

\vskip 0.5cm

\centerline{\bigbf Figure captions}

\vskip 0.5cm

\item{Fig.1}
The one-loop diagrams for $\Pi^0_{00}$ (a) and for $\Pi^0_{11}$ (b).
The solid line is the bare electron propagator and the wavy line
represents the gauge field propagator. These are the leading order
diagrams of $\Pi_{00}$ and $\Pi_{11}$ in the $1/N$ expansion.

\vskip 0.5cm

\item{Fig.2}
The diagram that corresponds to the one-loop correction to the
fermion self energy.
The solid line is the bare electron propagator and the wavy line
represents the gauge field propagator.

\vskip 0.5cm

\item{Fig.3}
The diagrams that correspond to the $(1/N)^0$th order contributions
to $\Pi_{11}$ in the $1/N$ expansion.

\vskip 0.5cm

\item{Fig.4}
The diagram that corresponds to the lowest order
vertex correction
$\Gamma_0 ({\bf k}, {\bf q}, i\omega, i\nu)$ or
$\Gamma_1 ({\bf k}, {\bf q}, i\omega, i\nu)$.

\vskip 0.5cm

\item{Fig.5}
(a) The non-vanishing diagram generated by
$\psi^{\dagger} a_{\mu} A^{\mu} \psi$ vertex.
(b) A typical vanishing diagram generated by
$\psi^{\dagger} a_{\mu} A^{\mu} \psi$ vertex.

\bye